\documentclass[superscriptaddress,aps,prb,twocolumn,showpacs,floatfix,altaffilletter, longbibliography]{revtex4-2}
\usepackage{graphicx}
\usepackage{grffile}
\usepackage{amsmath}
\usepackage{amssymb}
\usepackage{bm}
\usepackage{color}
\usepackage[dvipsnames]{xcolor}
\usepackage{amsmath,amssymb,amsfonts}
\usepackage{epsfig}
\usepackage{times}
\usepackage[colorlinks,bookmarks=false,citecolor=blue,linkcolor=blue,urlcolor=blue]{hyperref}
\usepackage{dsfont}
\usepackage{multirow}
\usepackage{mathrsfs}

\usepackage{bbold}

\usepackage[customcolors,shade]{hf-tikz}

\usepackage{colortbl}
\usepackage{changes}

\newcommand{\sro}[0]{Sr$_2$RuO$_4$}

\newcommand{\fref}[1]{Fig.~\ref{#1}}
\newcommand{\eref}[1]{Eq.~(\ref{#1})}
\newcommand{\sref}[1]{Sec.~\ref{#1}}
\newcommand{\aref}[1]{Appendix~\ref{#1}}

\usepackage{fixfoot}

\usepackage{mathtools}
\usepackage{pifont}

\begin{document}

\title{
Frequency-Dependent Superconducting States From the Two-Time Linear Response Theory:\\ Application To Sr$_2$RuO$_4$
}
\author{Olivier~Gingras\email{ogingras@flatironinstitute.org}}
\affiliation{Center for Computational Quantum Physics, Flatiron Institute, 162 Fifth Avenue, New York, New York 10010, USA}

\author{Antoine~Georges}
\affiliation{Collège de France, 11 place Marcelin Berthelot, 75005 Paris, France}
\affiliation{Center for Computational Quantum Physics, Flatiron Institute, 162 Fifth Avenue, New York, New York 10010, USA}
\affiliation{Centre de Physique Théorique, Ecole Polytechnique, CNRS, Institut Polytechnique de Paris, 91128 Palaiseau Cedex, France}
\affiliation{DQMP, Université de Genève, 24 quai Ernest Ansermet, CH-1211 Genève, Suisse}

\author{Olivier~Parcollet}
\affiliation{Center for Computational Quantum Physics, Flatiron Institute, 162 Fifth Avenue, New York, New York 10010, USA}
\affiliation{Université Paris-Saclay, CNRS, CEA, Institut de physique théorique, 91191, Gif-sur-Yvette, France}

\date{\today}

\begin{abstract}
    We investigate the possible superconducting instabilities 
    of strongly correlated electron materials using a generalization of 
    linear response theory to external pairing fields depending on frequency.
    We compute a pairing susceptibility depending on two times, allowing us to capture dynamical pairing and in particular odd-frequency solutions.
    We first benchmark this method on the attractive one-band Hubbard model and then consider the superconductivity of 
    strontium ruthenate Sr$_2$RuO$_4$ within single-site dynamical mean-field theory, hence 
    restricting ourselves to pairing states which are momentum independent in the orbital basis.
    The symmetry of the superconducting order parameter of this material is still debated, and local odd-frequency states have been proposed to explain some experimental discrepancies.
    In the temperature range studied, we find that the leading eigenvectors are odd-frequency intra-orbital spin-triplet states, while the eigenvectors with the highest predicted transition temperature correspond to even-frequency intra-orbital spin-singlet states. 
    The latter include a state with $d$-wave symmetry when expressed in the band basis.
\end{abstract}

\maketitle

\section{Introduction}

A number of materials with strong electronic correlations display unconventional superconductivity (SC). 
The term `\textit{unconventional}' is often used to imply that the SC is caused by electron-electron interactions rather 
than electron-phonon interactions, or that the superconducting order parameter (SCOP) has symmetry properties 
that differ from conventional $s$-wave. 
These two aspects are often connected: in a material with strong repulsive interactions between electrons, 
equal-time local pairing is strongly suppressed.  

One way to circumvent  the on-site repulsion is when the SCOP is spatially non-local, \textit{i.e.} momentum dependent, 
and that the integral of the gap function over momentum vanishes (as in the $d$-wave state of the cuprates for example~\cite{scalapino_case_1995, Gu2019}).
Another way is retardation: if the equal-time pairing amplitude vanishes, the electrons forming a pair follow 
each other with a time-delay and the local repulsion can be avoided. 
`\textit{Odd-frequency}' pairing, originally introduced by Berezinskii 
in the context of superfluid Helium 3~\cite{berezinskii_new_1974}, 
precisely enforces that the equal-time pairing amplitude vanishes due to the antisymmetry 
of the SCOP in the time-domain -- see \cite{linder_odd-frequency_2019, triola_annphys_2020} for reviews. 
In multi-orbital systems, an even larger set of possibilities exist for unconventional SC, making use of the non-trivial symmetry properties 
of the SCOP under spin, parity, orbital and time (retardation), as described by the $SPOT$ classification~\cite{linder_odd-frequency_2019, gingras_superconductivity_2022}.
In particular, multiple different $SPOT$ classes can coexist and it was shown that odd-frequency correlations are ubiquitous in multi-orbital superconducting systems~\cite{PhysRevB.88.104514, PhysRevB.92.094517, triola_annphys_2020, gingras_superconductivity_2022}.

From a computational standpoint, stabilizing the SC phase is often very challenging because it involves temperatures and 
energy scales that are much lower than the bare electronic energy scales. 
Instead, one can assess the tendency towards SC by performing computations at higher temperature and computing 
a pairing susceptibility, \textit{i.e.} the response of the system to an external pairing field. 
One way to do this is to treat the pairing field in perturbation theory in order to obtain the pairing susceptibility, 
\textit{i.e.} to construct the linearized Eliashberg equations and to find the dominant eigenvalues of the associated kernel~\cite{eliashberg1960interactions, nourafkan_correlation-enhanced_2016, gingras_superconducting_2019, gingras_superconductivity_2022}. 
However, this requires a computation of the two-particle susceptibilities and vertex functions which is a quite demanding task.    

Here, we present the two-time linear response theory (TTLR), a computational method that allows us to compute generalized pairing susceptibilities for unequal-time 
pairing by explicitly introducing a pairing forcing field and directly computing the response of the system. 
We implement this method in the framework of dynamical mean-field theory (DMFT)~\cite{RevModPhys.68.13}, 
hence limiting ourselves to spatially local pairing. 
The TTLR method is conceptually similar to the one introduced in Ref.~\citenum{Georges_1993}, 
but is improved and generalized by using an expansion on basis functions in the time domain 
(Legendre polynomials). 
Furthermore, we implement it using a continuous-time quantum Monte Carlo algorithm in the Nambu formalism~\cite{RevModPhys.83.349, SETH2016274}, 
allowing accurate solutions of the DMFT equations for temperatures down to $1/80~\text{eV}$.

We first demonstrate the usefulness of TTLR by applying it to the single-band Hubbard with an 
attractive interaction $U<0$. In this case, a solution of the DMFT equations is possible directly 
in the SC state, hence this serves as a benchmark. We obtain excellent agreement with the direct 
solution. 

We then consider the unconventional superconductor \sro \,(SRO). 
Owing to single crystals of extremely high quality, 
this compound is one of the most thoroughly studied among strongly correlated materials, with many different 
experimental probes~\cite{RevModPhys.75.657}. Its normal state displays large orbital-selective 
effective mass enhancements~\cite{bergemann_normal_2001}, and a crossover from a Fermi liquid 
below $T\sim 25$~K to a less coherent metal above that scale~\cite{hussey_normal-state_1998, PhysRevB.58.R10107, kugler_strongly_2020}. 
It is now understood that the normal state inherits its strong correlations from the combined effect 
of the Hund's rule coupling and the proximity of the $\gamma$-band to a 
van Hove singularity~\cite{mravlje_coherence-incoherence_2011}. 
Recent RIXS experiments~\cite{Suzuki_2023} have confirmed that SRO belongs to the broad family 
of Hund metals~\cite{Yin_Hund_2011,georges_strong_2013} 
and DMFT provides a quantitative description of many  
of its normal state properties~\cite{mravlje_seebeck_2016, tamai_high-resolution_2019, strand_magnetic_2019, PhysRevLett.131.236502, blesio2023signatures}. 

In contrast, the nature of the unconventional SC state of SRO, discovered 
almost thirty years ago~\cite{Maeno1994, RevModPhys.75.657}, is 
still being actively debated, together with the symmetry of its SCOP~\cite{mackenzie_even_2017}. 
Only recently was it experimentally shown to be dominantly composed of spin-singlet pairs~\cite{pustogow_constraints_2019, ishida_reduction_2020}.
The debates persist because thermodynamic measurements  suggest a one-component order parameter~\cite{hassinger_vertical_2017, Li2022, doi:10.1073/pnas.2020492118, PhysRevB.106.064513}, while other experiments observed evidence of a two-component order parameter~\cite{Benhabib2021, Ghosh2021, Grinenko2021}.
Recent computational efforts have been devoted to elucidating this question, starting from 
a first-principle description of the electronic structure of SRO, \textit{e.g.} 
from an analysis of the Eliashberg equations in combination 
with DMFT~\cite{gingras_superconducting_2019, Acharya2019, PhysRevB.105.155101, app11020508, PhysRevResearch.5.L022058} 
or using the functional renormalisation group~\cite{hauck2023competition}.
Considering some of these works found local and odd-frequency dominant eigenstates~\cite{gingras_superconducting_2019, PhysRevB.105.155101, gingras_superconductivity_2022} and argue they could help explain experimental discrepancies~\cite{gingras_superconductivity_2022}, in this work, we employ the TTLR method to study local unequal-time pairing states of SRO within the DMFT framework.

The paper is organized as follow. TTLR is presented in details in \sref{sec:method}.
In \sref{sec:hubbard}, we apply it to the attractive Hubbard model which serves as a benchmark. 
We discuss the convergence with respect to the number of Legendre polynomials in the basis 
and the conditions on the amplitudes of the pairing fields to remain in the linear response regime. 
We analyse the temperature dependence of the resulting eigenvectors and their frequency structure.
As shown in \aref{app:power}, we recover the solutions obtained using the power method.
In \sref{sec:sro}, we generalize the method to multi-orbital systems and apply it to SRO.
We find that, at high-temperatures, the eigenvectors with largest eigenvalues are spin-triplet intra-orbital odd-frequency states similar to those obtained in Refs~\citenum{gingras_superconductivity_2022} and \citenum{gingras_superconducting_2019}.
Moreover, the inter-orbital odd-frequency states reported in Ref.~\citenum{PhysRevB.105.155101} are found to be subdominant.
Finally, the states with the highest interpolated critical temperatures correspond to 
even-frequency spin-singlet states.
We show that, although local in momentum space in the orbital basis, the upfolding to the band basis of one of these states corresponds to $d$-wave pairing.

\section{Method}\label{sec:method}

In this section, we present our approach to study a general SC order 
using a time-dependent field coupled to the superconducting order (referred to as the "SC field" below)
in the linear regime, in imaginary time. 
This is a straightforward generalization of the standard linear response theory technique using a small {\it static} field.
However, this generalization allows us to study more general SC orders, and in particular the {\it odd-frequency pairing} phases.
After establishing our notations in \sref{sec:method:notations}, we show in \sref{sec:method:Leg} 
how to decompose the SC susceptibility on a compact basis of Legendre polynomials.
This approach is general and allows us to address both even- and odd-frequency pairing.

\subsection{Notations}\label{sec:method:notations}

Here, we first establish some notations, especially in relation to conventions on Nambu spinors. 
We then detail the application of the TTLR method within DMFT.

\subsubsection{Green's function in Nambu space.}\label{sec:method:notations:gfnambu}
We consider a system with a single orbital and two spin states per unit cell.
The generalization to more quantum numbers is given, in particular for
multi-orbital systems, in \sref{sec:sro}.  This system is described by an
Hamiltonian $\mathcal{H} = \mathcal{H}_0 + \mathcal{H}_{\text{int}}$, where 
$\mathcal{H}_0$ is the non-interacting quadratic part  and $\mathcal{H}_{\text{int}}$ is the interacting part.  
Here we neglect spin-orbit coupling, making
$\mathcal{H}_0$ spin-diagonal and $SU(2)$ symmetric, and
$\mathcal{H}_{\text{int}}$ is taken to be local.

The non-interacting Hamiltonian $\mathcal{H}_0$ can be Fourier transformed to momentum space and reads
\begin{equation}
    \label{eq:nambu_non-int_ham}
    \mathcal{H}_0 = \sum_{\textbf{k}\sigma} \epsilon_{\textbf{k}} \psi^{\dag}_{\textbf{k}\sigma} \psi_{\textbf{k}\sigma}
\end{equation}
where $\epsilon_{\textbf{k}}$ is the spin-independent band dispersion
and the operators $\psi_{\textbf{k}\sigma}$ and $\psi^\dag_{\textbf{k}\sigma}$ are destruction and creation operators of an electron with momentum $\textbf{k}$ and spin $\sigma$.
Defining the Nambu spinors
\begin{equation}
    \label{eq:nambu_spinor}
    \hat{\Psi}^{\dag}_{\textbf{k}}
        \equiv \left( \psi^{\dag}_{\textbf{k} \uparrow} \quad \psi_{-\textbf{k}\downarrow}\right),
\end{equation}
this Hamiltonian reads (up to a constant)
\begin{equation}
    \mathcal{H}_0 = \sum_{\textbf{k}} \hat{\Psi}^{\dag}_{\textbf{k}} \hat{\mathcal{H}}_{0\textbf{k}} \hat{\Psi}_{\textbf{k}}
    \ \ \text{with} \ \
    \hat{\mathcal{H}}_{0\textbf{k}} = \left[
        \begin{array}{cc}
            \epsilon_{\textbf{k}} & 0 \\
            0 & - \epsilon_{-\textbf{k}}
        \end{array}
    \right].
\end{equation}
In the following, the hat symbol indicates that the object is expressed in the Nambu basis.
Similarly, the number of electron operator $N$ can be expressed as
\begin{equation}
    \label{eq:nambu_number_op}
    N = \sum_{\textbf{k}} \hat{\Psi}_{\textbf{k}}^\dag \hat{\tau_3} \hat{\Psi}_{\textbf{k}},
\end{equation}
with $\hat{\tau}_3$ the third Pauli matrix. 

The partition function $Z_{\phi}$ is given by
\begin{equation}
    Z_{\phi} = \int \mathcal{D} \Psi \mathcal{D} \Psi^\dag e^{-S_0[\Psi, \Psi^{\dag}] -S_U[\Psi, \Psi^{\dag}] -S_{\phi}[\Psi, \Psi^{\dag}]}
\end{equation}
where the action is split in three contributions: the non-interacting part, the interaction part and the source field driven part, respectively given by
\begin{align}\label{eq:action}
    %
   S_{0}[\Psi, \Psi^\dag] & \equiv \sum_{\textbf{k}} \int_0^{\beta} d\tau \ \hat{\Psi}^{\dag}_{\textbf{k}}(\tau)\left(\hat{\partial}_\tau + \hat{\mathcal{H}}_{0\textbf{k}} - \mu \hat{\tau}_3 \right)\hat\Psi_{\textbf{k}}(\tau), \nonumber \\
    S_{U}[\Psi, \Psi^\dag] & \equiv \sum_{\textbf{k}} \int_0^{\beta} d\tau \ \mathcal{H}_{\text{int}}[\Psi, \Psi^\dag], \\  
    S_{\phi}[\Psi, \Psi^\dag] & \equiv - \sum_{\textbf{k}} \int_0^{\beta} d\tau d\tau' \ \hat{\Psi}^{\dag}_{\textbf{k}}(\tau)\hat{\phi}_{\textbf{k}}(\tau-\tau')\hat\Psi_{\textbf{k}}(\tau'), \nonumber
\end{align}
with the chemical potential denoted by $\mu$ and the dynamical source field by $\hat{\phi}_{\textbf{k}}(\tau)$. The inverse temperature $\beta$ is expressed in eV$^{-1}$ throughout the text.
In the Nambu basis, this field is expressed as the particle-hole ($ph$) and particle-particle ($pp$) components as
\begin{equation}
    \label{eq:phi_pp_ph}
    \hat{\phi}_{\textbf{k}} = \left( \begin{array}{cc} 
        \phi^{ph}_{\textbf{k}} & \phi^{pp}_{\textbf{k}} \\ 
        \phi^{\bar{pp}}_{\textbf{k}} & \phi^{\bar{ph}}_{\textbf{k}}
    \end{array} \right).
\end{equation}

The free-energy $\mathcal{F}_\phi$ in the presence of $\phi$ is related to the partition function by
$\mathcal{F}_{\phi} \equiv -\frac{1}{\beta} \ln Z_{\phi}$ and the corresponding Nambu Green's function is
\begin{align}
    \label{eq:nambu_green}
    \hat{G}_{\phi;\textbf{k}ab}(\tau-\tau') 
    & = \frac{\delta \mathcal{F}_{\phi}}{\delta  \hat{\phi}_{\textbf{k}ba}(\tau'-\tau)} \\
    \label{eq:nambu_green2}
    & = - \Big \langle T_{\tau} \hat \Psi_{\textbf{k}a}(\tau-\tau') \hat \Psi^{\dag}_{\textbf{k}b}\Big\rangle_{\phi}
\end{align}
where $a,b$ are Nambu indices, the imaginary-time dependence is given in the interaction picture as $A(\tau) \equiv e^{\tau (\mathcal{H} - \mu N)} A e^{-\tau (\mathcal{H} - \mu N)}$ and
\begin{equation}
    \langle A \rangle_{\phi} \equiv \frac{1}{Z_{\phi}}\int \mathcal{D}\Psi \mathcal{D} \Psi^\dag \ e^{-S_0-S_U-S_\phi} A.
\end{equation}
Using matrix notations in Nambu space for $G$, we have
\begin{align}
    \hat{G}_{\phi; \textbf{k}}(\tau) 
        & = -\left(
        \begin{array}{cc}
            \langle T_{\tau} \psi_{\textbf{k}\uparrow}(\tau) \psi^{\dag}_{\textbf{k}\uparrow}\rangle_{\phi} &
            \langle T_{\tau} \psi_{\textbf{k}\uparrow}(\tau) \psi_{-\textbf{k}\downarrow}\rangle_{\phi} \\
            \langle T_{\tau} \psi^{\dag}_{-\textbf{k}\downarrow}(\tau) \psi^{\dag}_{\textbf{k}\uparrow}\rangle_{\phi} &
            \langle T_{\tau} \psi^{\dag}_{-\textbf{k}\downarrow}(\tau) \psi_{-\textbf{k}\downarrow}\rangle_{\phi}
        \end{array} \right) \nonumber \\
        & \quad \quad \equiv
        \left(
        \begin{array}{rr}
            G_{\phi; \textbf{k}}(\tau) & F_{\phi; \textbf{k}}(\tau) \\
            \bar{F}_{\phi; \textbf{k}}(\tau) & \bar{G}_{\phi; \textbf{k}}(\tau)
        \end{array} \right)
\end{align}
where $G$ and $\bar{G}$ are respectively the particle and hole propagators or normal Green's functions, while $F$ and $\bar{F}$ are the  anomalous Green's functions.
The Dyson equation for the Green's function \eref{eq:nambu_green2} reads
\begin{equation}
    \label{eq:eq_motion_nambu_field}
    \hat{G}_{\phi; \textbf{k}} (i\omega_n)^{-1} = i\omega_n \hat{\tau}_0 + \mu \hat\tau_3 - \hat{\mathcal{H}}_{0\textbf{k}} - \hat{\phi}_{\textbf{k}}(i\omega_n) - \hat \Sigma_{\phi; \textbf{k}}(i\omega_n)
\end{equation}
where $\hat{\tau}_0$ is the identity and $\hat \Sigma$ is the self-energy in the Nambu basis.
In what follows, we are considering only pairing fields, taken to be purely local.
That is $\phi^{ph} = \phi^{\bar{ph}} = 0$ and $\hat{\phi}_{\textbf{k}} = \hat{\phi}$.

Finally, in linear response to $\phi$, the definition of the superconducting susceptibility $\chi^{pp}$ given in \eref{eq:pp_suscep} can be inverted to find the anomalous Green's function $F$ in the presence of a field as
\begin{equation}
    \label{eq:lin_resp}
    F_\phi(\tau) = \int d\tau' \ \chi^{pp}(\tau, \tau') \phi(\tau') + \mathcal{O}(\phi^2).
\end{equation}

As shown in \aref{sec:sym_F}, even-frequency superconductivity corresponds to the symmetry property 
$F(\tau) = -F(\beta-\tau)$ while odd-frequency corresponds to $F(\tau) = F(\beta-\tau)$.
According to the $SPOT$ classification~\citenum{linder_odd-frequency_2019} detailed in App.~\ref{sec:SPOT}, 
in the single orbital context with local pairing, even-frequency must correspond to singlet pairing and 
odd-frequency to triplet pairing.
In the multi-orbital case, there are additional possibilities, which will be reviewed in 
Sec.~\ref{sec:sro}.


\subsubsection{\label{sec:method_dmft}Dynamical mean-field theory.}

We now rewrite explicitly the DMFT equations in Nambu space~\cite{RevModPhys.68.13} in the presence of time-dependent source fields.

Given a dynamical Weiss field $\hat{\mathcal{G}}_{\phi}^{0}(i\omega_n)^{-1}$ which specifies the quadratic part 
of the local action, the effective impurity model is solved, in the presence of the source fields, to obtain 
the impurity Green's function:
\begin{equation}
    \hat{G}_{\phi}^{\text{imp}}(i\omega_n) = - \int_0^\beta d\tau e^{i\omega_n \tau} \langle T_{\tau} \hat{\Psi}_i(\tau) \hat{\Psi}_i^\dag \rangle_{S_{\phi}^{\text{eff}}}
\end{equation}
with $\hat{\Psi}^{\dag}_i = \left( \psi^{\dag}_{i \uparrow} \quad \psi_{i\downarrow}\right)$. 
This also yields the impurity self-energy: 
\begin{equation}
    \hat{\Sigma}_{\phi}^{\text{imp}}(i\omega_n) = \hat{\mathcal{G}}_\phi^0(i\omega_n)^{-1} - \hat{G}_\phi^{\text{imp}}(i\omega_n)^{-1}.
\end{equation}
This local self-energy is then used in the lattice Dyson equation to construct the local component of the 
lattice Green's function as:
\begin{equation}
    \hat G_{\phi}^{\text{loc}}(i\omega_n) = \sum_{\textbf{k}} \hat G_{\phi; \textbf{k}}(i\omega_n).
\end{equation}
and the Weiss field is then updated according to:
\begin{equation}
    \label{eq:weiss_latt}
    \hat{\mathcal{G}}_{\phi}^{0}(i\omega_n)^{-1} = \hat{\Sigma}_{\phi}(i\omega_n) + \hat{G}_{\phi}^{\text{loc}}(i\omega_n)^{-1},
\end{equation}

The DMFT calculations presented below are performed using the TRIQS
package~\cite{PARCOLLET2015398} and the impurity problem is solved using the
continuous-time quantum Monte Carlo algorithm in the hybridization-expansion
formulation (CT-HYB)~\cite{PhysRevLett.97.076405, PhysRevB.75.155113, RevModPhys.83.349, SETH2016274}. 
We use the
formulation in which the imaginary-time Green's function is purely real and for
which the hermiticity condition is enforced, leading to 
\begin{align}\label{eq:F_Fbar} 
   F_\phi^{\text{imp}}(\tau) = \bar{F}_\phi^{\text{imp}}(\tau).
\end{align} 
In what follows, we use $F \equiv F^{\text{imp}}$.
Similarly, $\phi \equiv \phi^{pp} = \phi^{\bar{pp}}$.

\subsection{\label{sec:method:Leg}Legendre representation of the superconducting susceptibility}

We now discuss the external pairing fields $\phi(\tau)$ that depend on imaginary time.
Using a full basis of fields, we will have a complete determination of 
the dynamical pairing susceptibility $\chi^{pp}$, from which we will get the 
superconducting eigenvectors, including their full frequency dependence.
In practice, we need a computation for each basis field, which generates a pairing response $F_{\phi}(\tau)$.
It is therefore crucial to use a compact basis for the time dependency of the field.

We use the basis of Legendre polynomials normalized and scaled to imaginary-time, truncated at the $N$th polynomial. This basis provides a compact representation of the Green’s functions~\cite{PhysRevB.84.075145}.
We denote it by $\{P_\alpha(\tau)\}_{\alpha\in\{0,...,N-1\}}$, with explicit form and properties detailed in \aref{sec:leg}.
Note that the polynomials are rescaled and shifted to represent a function on $[0,\beta]$.
Using this basis, a general external field reads, for $0 \leq \tau \leq \beta$:
\begin{equation}
    \label{eq:general_phi}
    \phi(\tau) = \sum_{\alpha} \phi_\alpha P_\alpha(\tau)
\end{equation}
where the $\phi_\alpha$ are the expansion coefficients.
In the linear regime with \eref{eq:lin_resp}, we get
\begin{align}
    \label{eq:expansion_F}
    F_{\phi}(\tau) & = \sum_\alpha \phi_\alpha \chi_\alpha(\tau) + \mathcal{O}(\phi^2), \quad \text{where} \nonumber \\   
    \chi_\alpha(\tau) & \equiv \int d\tau' \chi^{pp}(\tau, \tau') P_{\alpha}(\tau').
 \end{align}
Here, $\chi_\alpha(\tau)$ is the susceptibility in a mixed Legendre-time basis.
It can be straightforwardly rewritten fully in the Legendre basis with coefficients  $\chi_{\alpha}^{\beta}$ as
\begin{equation}
    \chi_\alpha(\tau) = \sum_{\beta} \chi_\alpha^{\beta} P_\beta(\tau).
\end{equation}

The coefficient $\chi_\alpha^\beta$ can be extracted using the orthogonality relation of the Legendre polynomials discussed in \aref{sec:leg}, that is
\begin{equation}
    \frac{2}{\beta} \int_0^\beta d\tau P_\alpha(\tau) P_\beta(\tau) = \delta_{\alpha\beta}.
\end{equation}
Then,
\begin{align}
    \chi_{\alpha}^{\beta} & = \frac{2}{\beta} \int_0^\beta d\tau P_\beta(\tau) \chi_{\alpha}(\tau).
\end{align}

We emphasize that, in $\chi_\alpha^\beta$, the subscript denotes the Legendre component of the external field used to generate the dynamical pairing response.
The superscript denotes the projection of this component on the Legendre basis.
We also note that the polynomials $P_\alpha(\tau)$ with even $\alpha \in 2\mathbb{N}$ are even around $\beta/2$ are correspond to odd-frequency states, while odd $\alpha \in
2\mathbb{N}+1$ are odd around $\beta/2$ and correspond to even-frequency states, as discussed in \aref{sec:sym_beta_2}.
We show in \aref{sec:sym_F}
that dynamical pairing responses have the same parity around $\beta/2$ as the fields used to generate them.
In other words, the $\chi_{\alpha}^{\beta}$ susceptibility
is block diagonal in these even and odd indices.  
From now on, when discussing linear combinations of
Legendre polynomials that are even (odd) around $\beta/2$, we will refer to
them as odd-frequency (even-frequency).


Once the $N\times N$ matrix $\chi_{\alpha}^{\beta}$ is constructed, the leading superconducting eigenvectors are obtained by diagonalizing
it.
Its eigenvectors $\varphi_l(\tau)$ with eigenvalues $\lambda_l$ satisfy
\begin{equation}
    \lambda_l \varphi_l(\tau) = \int d\tau' \chi^{pp}(\tau, \tau') \varphi_l(\tau').
\end{equation}
The convergence as a function of the size of the Legendre basis $N$ is studied in explicit cases
in the next sections.

Finally, let us discuss in this context the simple method of using a {\it static} 
field.
A static superconducting field reads in the current language (due to the anti-periodicity condition):
\begin{equation}
    \phi_{\text{static}}(\tau) = \phi_0 \left( \delta(\tau) - \delta(\beta-\tau) \right).
\end{equation}
It can be checked that this expression produces an equaltime term in \eref{eq:action}.
A static field computation is sufficient to find the superconducting transition, 
if $\phi_{\text{static}}$ is not orthogonal to the eigenvector corresponding to the diverging 
eigenvalue of the $\chi$ matrix.
However, the odd-frequency solutions occur in a different sector, which {\it is} orthogonal 
to  $\phi_{\text{static}}$, as $ \phi_{\text{static}}$ is orthogonal to each Legendre of even degree (as $P_{2\alpha}(0) = P_{2\alpha}(\beta)$).
Therefore they are not accessible with such a simple technique and require the full dynamical calculation presented above.

\section{\label{sec:hubbard}Benchmark on the attractive Hubbard model}

First, we benchmark our method on a simple case: the one-band attractive ($U<0$) Hubbard model 
on the Bethe lattice at half-filling~\cite{PhysRevB.69.184501, PhysRevLett.86.4612, Toschi_2005, PhysRevA.84.023638}.
The model displays a crossover between the weak coupling Slater regime in which the BCS mechanism applies, and the strong coupling Mott/Heisenberg regime in which the superconducting transition at $T_c\propto t^2/|U|$ corresponds to the Bose condensation of pairs that form at a much higher temperature $\sim |U|$~\cite{PhysRevB.44.5190, PhysRevB.54.13138, PhysRevB.54.1286, Pedersen1997, PhysRevB.72.235118, Gersch_2008}.
An exact mapping is known between the superconducting phase of the $U<0$ model and the antiferromagnetic solution of the repulsive case $U>0$~\cite{PhysRevA.79.033620}.

The Hubbard Hamiltonian on the Bethe lattice is given by $\mathcal{H} = \mathcal{H}_0 + \mathcal{H}_{\text{int}}$ with
\begin{equation}
    \mathcal{H}_0 = -\,\frac{t}{\sqrt{z}} \sum_{\langle i, j\rangle \sigma} \left[\psi^\dag_{i\sigma} \psi_{j\sigma} + \psi^\dag_{j\sigma} \psi_{i\sigma} \right] - \mu \sum_{i\sigma} n_{i\sigma}
\end{equation}
where $z$ is the connectivity of the lattice ($z\rightarrow\infty$), 
$t$ the nearest neighbor hopping amplitude, $\langle i,j \rangle$ denotes pairs of nearest neighbors, $\mu = U/2$ is the chemical potential at half-filling, $\psi_{i\sigma}$ ($\psi^\dagger_{i\sigma}$) is the annihilation (creation) operator of an electron on site $i$ with spin $\sigma$ and $n_{i\sigma} = \psi^\dagger_{i\sigma}\psi_{i\sigma}$, and
\begin{equation}
    \mathcal{H}_{\text{int}} = U \sum_i n_{i\uparrow}  n_{i\downarrow}
\end{equation}
where $U$ is the on-site Coulomb repulsion.

On the Bethe lattice, the relation between the DMFT Weiss field $\hat{\mathcal{G}}_0$ and the lattice Green's function $\hat{G}$ of \eref{eq:weiss_latt} is exactly given by
 \begin{equation}
     \hat{\mathcal{G}}^0_\phi(i\omega_n)^{-1} = i\omega_n \hat{\tau}_0 + \mu \hat{\tau}_3 - t^2\hat{\tau}_3 \hat{G}_\phi(i\omega_n) \hat{\tau}_3 - \hat{\phi}(i\omega_n).
 \end{equation}
We consider only the case with $t=1$ and $U=-3$, known to have a superconducting transition around $\beta \sim 5.5$~\cite{PhysRevA.84.023638}.

\begin{figure}[t]
    \centering
    \includegraphics[width=\linewidth]{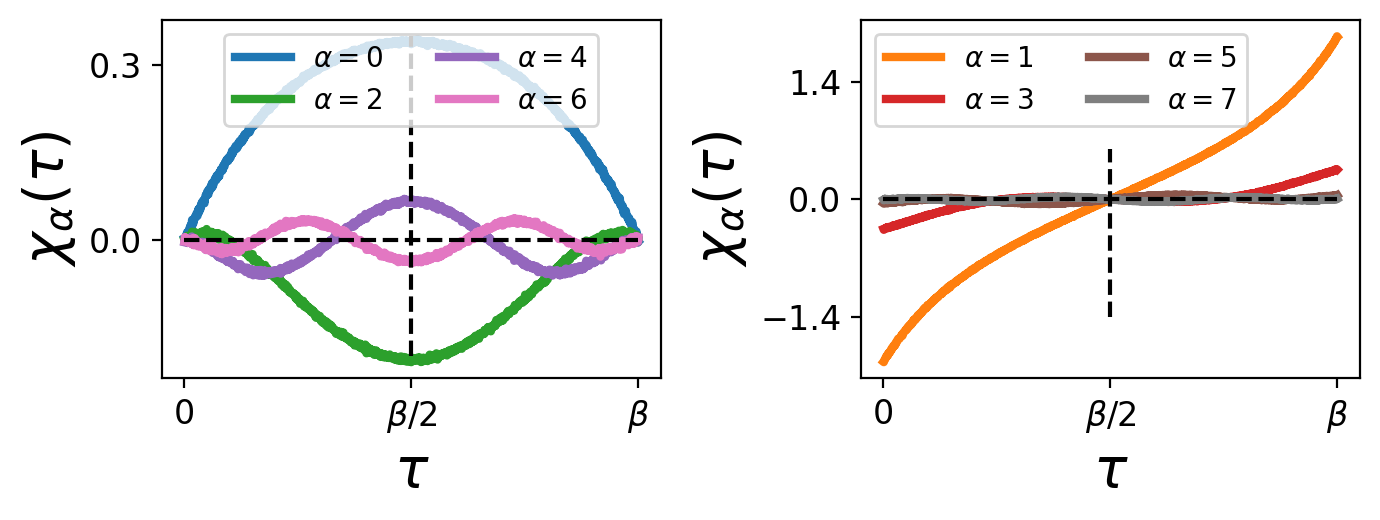}
    \caption{
    Dynamical pairing responses to external pairing fields for the half-filled attractive Hubbard model with $U=-3t$.
    The fields are proportional to Legendre polynomials $P_\alpha(\tau)$ that generate the purely a) odd-frequency and b) even-frequency finite dynamical pairing responses $\chi_\alpha(\tau)$.
    We used $t=1$, $\beta = 3$ and $\phi_\alpha= 0.01 \ \forall \ \alpha$.
    }
    \label{fig:att_response_beta3}
\end{figure}

\begin{figure}[b]
    \centering
    \includegraphics[width=\linewidth]{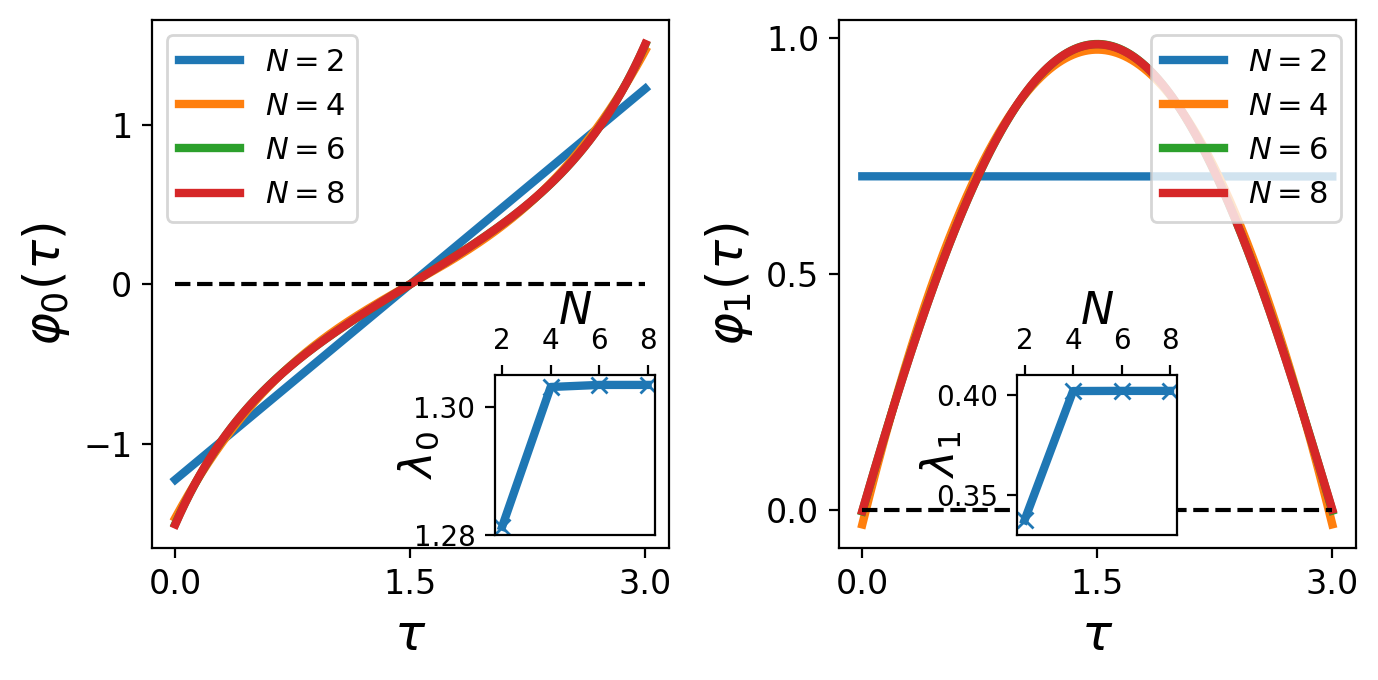}
    \caption{
    Convergence of the a) leading and b) subleading eigenvectors with respect to the size of the Legendre basis $N$.
    The system is the half-filled attractive Hubbard model on the Bethe lattice with the same parameters used as in \fref{fig:att_response_beta3}.
    Inset: convergence of the associated eigenvalue versus $N$.
    }
    \label{fig:att_eigvec_beta3}
\end{figure}

We compute the response of the system to a set of pairing fields by converging the DMFT equations in the 
presence of these fields, which have an imaginary-time structure given by Legendre polynomials. 
In practice, we first converge the DMFT equations in the normal state in the absence of pairing fields, 
then turn on the pairing fields and proceed with a few additional iterations until convergence is reached.
Figure~\ref{fig:att_response_beta3}~(a) shows the dynamical pairing responses in imaginary-time $\chi_{\alpha}(\tau)$  resulting from the odd-frequency external pairing fields $\phi(\tau) = \phi_{\alpha} T_{\alpha}(\tau)$.
Figure~\ref{fig:att_response_beta3}~(b) shows the responses to even-frequency fields.
As expected for the case presented in this figure, the responses tend to be larger for even-frequency superconductivity than for odd-frequency.

We use the resulting pairing responses to construct the susceptibility matrix $\chi_{\alpha}^{\beta}$.
The eigenvectors with largest and second largest eigenvalues are shown in \fref{fig:att_eigvec_beta3}~(a) and (b) respectively, as a function of the basis size $N$, again for $\beta=3$.
The imaginary-time structure seems converged around $N\sim 4$, confirmed by the inset that presents the evolution of the eigenvalue as a function of $N$.

\begin{figure}[t]
    \centering
    \includegraphics[width=\linewidth]{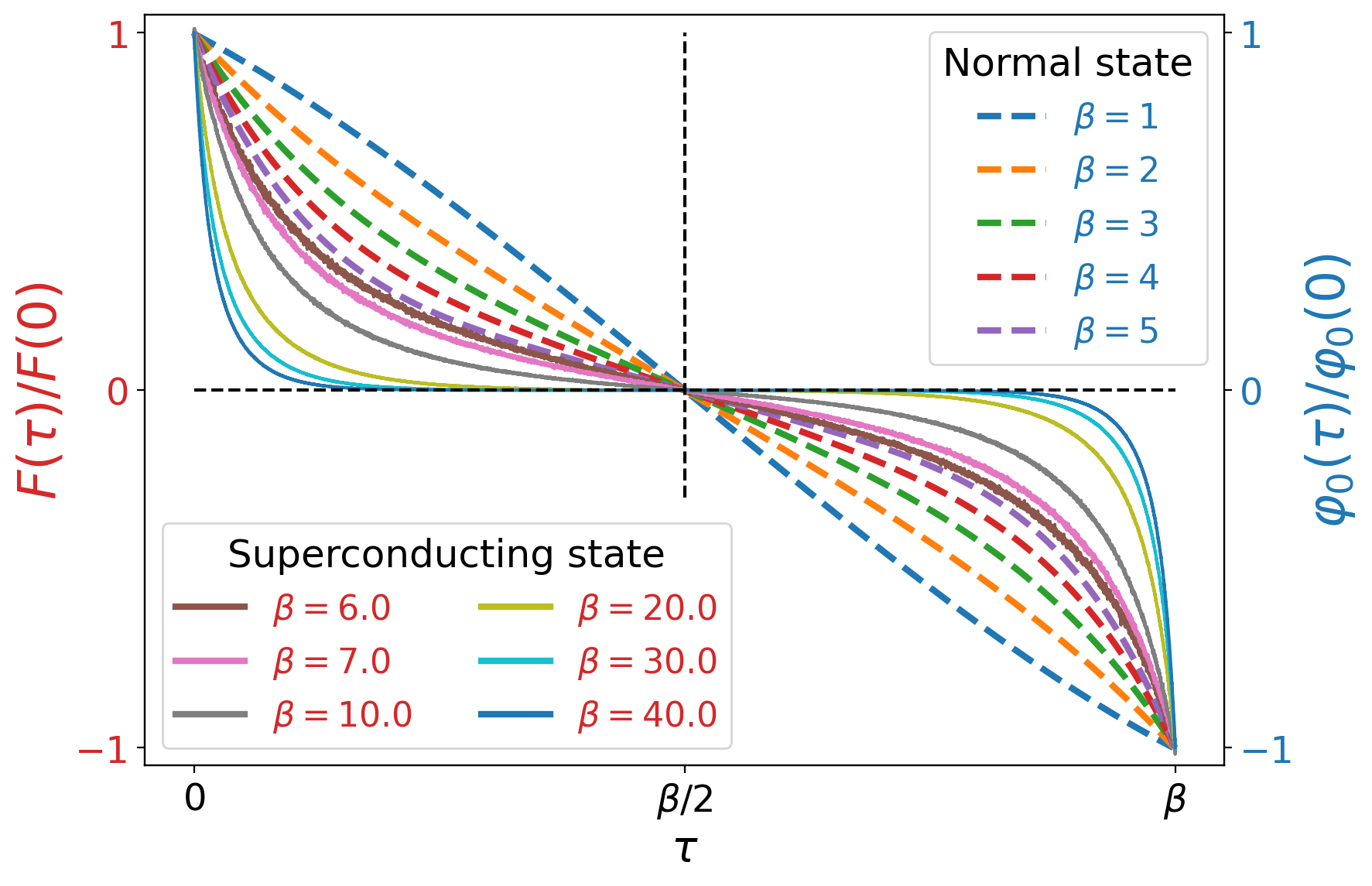}
    \caption{
    Imaginary-time-structure evolution of the leading superconducting eigenvector (dash lines, right axis) into the anomalous Green's function (solid line, left axis) across the superconducting transition in the half-filled attractive Hubbard model.
    The same parameters as in \fref{fig:att_response_beta3} were used and the leading eigenvectors were obtained with a Legendre basis of size $N=8$.
    }
    \label{fig:att_hub_U-3}
\end{figure}

We repeat this process for different temperatures while making sure to remain in the linear regime, as discussed in \aref{sec:lin_regime}.
In \fref{fig:att_hub_U-3}, we present the evolution of the leading superconducting eigenvector $\varphi_0$ with respect to inverse temperature $\beta$.
The dashed lines with blue labels correspond to the leading eigenvectors from the normal state obtained with the dynamical susceptibility.
The solid lines with red labels correspond to the actual anomalous Green's function $F(\tau)$, which naturally develops once the system has transitioned to the superconducting state.
All these states are normalized to have $\varphi_0(\tau = \beta) = 1$ and $F(\tau = \beta)=1$.

From this figure, we see clearly that at low temperature, most of the pairing happens at equal-time, which can be well captured by the static field method.
Around the critical temperature however, the eigenvectors are much more time-dependent, which is well captured by the eigenvectors of the dynamical pairing susceptibility.

\begin{figure}[b]
    \centering
    \includegraphics[width=.8\linewidth]{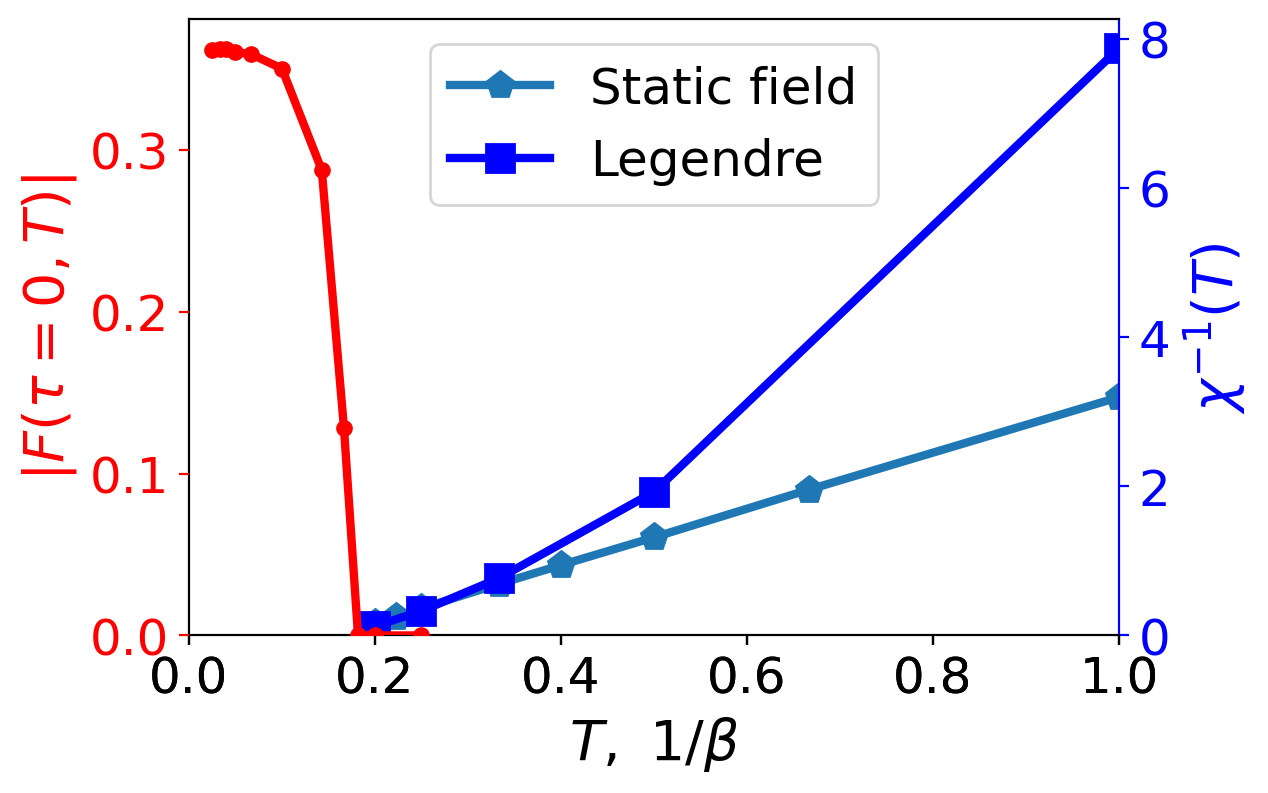}
    \caption{
    The superconducting critical temperature in the half-filled attractive Hubbard model  for $U/t=-3$. 
    The red left axis presents the spontaneous appearance of the pairing amplitude in the superconducting state around $T\sim 1/6$. 
    The blue right axis shows that this critical temperature is indicated from the normal state by a divergence of the dynamical pairing susceptibility. 
    We compare two methods to obtain the inverse susceptibility: dark squares for the two-time linear response and light blue pentagons for the static method (in arbitrary units).
    }
    \label{fig:att_eigval_vs_T}
\end{figure}

Finally, we study the dependence on temperature of the eigenvalues of the dynamical pairing susceptibility 
and verify that the method correctly predicts the superconducting critical temperature.
We show in \sref{app:power} that the power method gives precisely the same results as those obtained using the dynamical pairing susceptibility.
The transition is signaled by the first diverging eigenvalue $\lambda_0$ of the susceptibility $\bm \chi^{pp}$ or equivalently by $1/\lambda_0 = 0$.
In \fref{fig:att_eigval_vs_T}, we plot $1/\lambda_0$ 
in the normal state obtained with both the static and dynamic methods (right axis, blue symbols), 
along with the superconducting dome that develops at low temperature (left axis, red symbols).

The darker blue squares are associated to the inverse of the leading eigenvalue of the dynamical pairing susceptibility.
The light blue pentagons represent the results using the static field method discussed in \sref{sec:method:Leg} 
and represent the inverse of $F(\tau=0)$ divided by the amplitude of the static field, which is given in arbitrary units.
After rescaling, we compare it to $1/\lambda_0$ obtained from the TTRL dynamical susceptibility.
This plot demonstrates the excellent agreement between the dynamical susceptibility and static field methods 
in determining the transition temperature.

The red curve shows the value of the anomalous Green's function at $\tau = 0$.
The red dome highlights the build-up of superconductivity at $\beta \geq 6$, which starts at the very same temperature predicted from the normal state.
Thus, the Legendre method succeeds to obtain the temperature of the superconducting transition for the single-band attractive Hubbard model at half-filling.
At the same time, it gives access to the frequency structure of many eigenvectors simultaneously, deep in the normal state.
For this model, as expected, it predicts an even-frequency state.


In conclusion for this section, we employed the two-time linear response theory to compute the dynamical pairing susceptibility in the intermediate coupling regime of the attractive Hubbard model.
In this regime, superconductivity appears at rather high temperature and is strongly dependent on imaginary-time.
We showed that, from the normal state, the leading eigenvalue of the dynamical pairing susceptibility provides information about not only the superconducting critical temperature, but also the superconducting order parameter.

\section{\label{sec:sro}Strontium ruthenate}
In this section, we apply our method to strontium ruthenate (SRO) within the DMFT framework.
The symmetry of the superconducting order parameter of SRO has been a mystery 
for almost 30 years now.
Previous works that studied the superconducting instabilities from the normal state by solving the Eliashberg equation in a formalism that allowed for frequency dependent SCOPs found local odd-frequency states as dominant candidates~\cite{gingras_superconductivity_2022, gingras_superconducting_2019, Acharya2019, PhysRevB.105.155101}.
The method detailed in this manuscript is designed to investigate such states 
and, more broadly in the case where it is employed in the DMFT framework, any pairing state in which the superconducting order parameter is local in space 
(\textit{i.e.} $\mathbf{k}$-independent in orbital space).  
Of course, states with strong momentum dependence such as $d$-wave pairing are strong contenders for 
this material.
While most of these states require an extension of our method 
to non-local generalizations of DMFT,
we show that for certain local gap functions in the orbital basis, a non-trivial momentum dependence such as $d$-wave can be obtained when upfolding to the band basis.

In \sref{sec:sro_multi-orb_model}, we present the minimal model that we use to 
describe SRO, which is downfolded from a density functional theory calculation.
We then generalize the formalism of \sref{sec:method} to multi-orbital systems.
We present a few dynamical pairing responses that highlight the coupling between different orbital sectors generated by the interaction.
In \sref{sec:sro_eigvecs}, we present the temperature dependence of the largest eigenvalues.
At every temperature considered, the leading instabilities are all intra-orbital odd-frequency spin-triplets, but we find that the states for which the extrapolated critical temperature is largest are intra-orbital even-frequency spin-singlet states.
Inspecting the frequency structure of these eigenvectors highlights that they take advantage of the strong retardation of the pairing to generate retarded pairing. 
In particular, one of these states transforms as a $d_{x^2-y^2}$ state once upfolded to the band basis.
We further compare these results with the literature, discuss them in the context of experiments and propose potential improvements for future works.

\subsection{\label{sec:sro_multi-orb_model}Minimal model and multi-orbital generalization of the method}

The main physics of SRO emerges from the metal-oxide planes, making it quasi-two-dimensional~\cite{bergemann_normal_2001, RevModPhys.75.657}.
While the strontium atoms simply act as reservoirs of electrons, the oxygen atoms (O) generate an octahedral structure around the ruthenium atom (Ru), acting as a crystal field.
This crystal field is large enough to split the partially filled $4d$ electronic shell of the Ru into four unoccupied $e_g$ and six $t_{2g}$ orbitals that effectively host four electrons.
Consequently, a minimal model for SRO can be constructed by retaining only the electronic states associated with the frontier
Wannier orbitals associated with Ru-O hybridized states of $t_{2g}$ character.

These orbitals are rather localized so that the electrons occupying them experience strong electronic correlations.
These correlations were shown to be essentially generated by local interactions of the Kanamori-Slater type, characterized by an on-site repulsion $U$ and Hund's coupling $J$~\cite{georges_strong_2013}.
Many experimental observations were 
well reproduced by DMFT computations
involving this minimal model and interactions, such as 
the magnitude and orbital selectivity of the effective 
mass enhancements~\cite{bergemann_normal_2001, mravlje_coherence-incoherence_2011}, 
the NMR response~\cite{mravlje_coherence-incoherence_2011}, 
the bad metal to Fermi liquid crossover~\cite{hussey_normal-state_1998, kugler_strongly_2020}, 
the Seebeck coefficient~\cite{mravlje_seebeck_2016}, 
the Fermi surface~\cite{tamai_high-resolution_2019}, 
the momentum-dependent spin response function~\cite{strand_magnetic_2019}
and the Raman response~\cite{blesio2023signatures}. 
The Hund's coupling was shown to play a crucial role~\cite{mravlje_coherence-incoherence_2011}, 
placing SRO among the broad family of 
Hund metals~\cite{Yin_Hund_2011,georges_strong_2013}. 

Spin-orbit coupling also plays a crucial role for 
SRO~\cite{PhysRevLett.85.5194, PhysRevLett.101.026406, PhysRevB.74.035115, PhysRevLett.112.127002, zhang_fermi_2016, Schaffer_2016, kim_spin-orbit_2018, tamai_high-resolution_2019}. 
However, including spin-orbit coupling in DMFT impurity solvers 
based on quantum Monte Carlo generates a large sign problem 
and is a challenging and active field of research.
For this reason, spin-orbit coupling is not included in the present work when solving the DMFT equations.

The downfolded Hamiltonian for the $t_{2g}$ Wannier functions~\cite{PhysRev.136.B864, PhysRev.140.A1133, Kohn1996} 
is obtained from a DFT calculation
~\cite{PhysRev.136.B864, PhysRev.140.A1133, Kohn1996} using the PBE exchange-correlation functional~\cite{Perdew:1996} 
with the Quantum Espresso package~\cite{giannozzi2009quantum, zhang2017p}.
The downfolding is performed by projecting the electronic wave function onto the set of maximally localized Wannier orbitals constructed using Wannier90~\cite{mostofi2008wannier90,mostofi2014updated,Pizzi_2020}.
Before applying external pairing fields, we converge the DMFT solution with $U=2.3$~eV and $J=0.4$~eV.
The DMFT calculations with and without external pairing fields were performed using the TRIQS package~\cite{PARCOLLET2015398} and the continuous-time quantum Monte Carlo algorithm~\cite{PhysRevLett.97.076405, RevModPhys.83.349, SETH2016274}.
The interface between Wannier90 and TRIQS is facilitated by the DFTTools package~\cite{AICHHORN2016200}.

Now, starting from a converged DMFT solution of SRO at a given temperature, we can compute the dynamical pairing susceptibility $\chi^{pp}$ using the formalism of \sref{sec:method}.
However, we need to generalize it for multi-orbital systems, in this case the three $t_{2g}$ orbitals labeled by $zx$, $yz$ and $xy$.
The Nambu spinor \eref{eq:nambu_spinor} becomes
\begin{align}
    \label{eq:sro_nambu_spinor}
    \hat{\Psi}^\dag_{\textbf{k}}
        & \equiv \left( \bm \psi^{\dag}_{\textbf{k} \uparrow} \quad \bm \psi_{-\textbf{k}\downarrow}\right) \quad \text{with} \\
    \bm \psi_{\textbf{k} \uparrow} 
         \equiv \left( \begin{array}{c}
            \psi_{\textbf{k} {yz} \uparrow} \\
            \psi_{\textbf{k} {zx} \uparrow} \\
            \psi_{\textbf{k} {xy} \uparrow}
        \end{array} \right) & \quad \text{and} \quad
     \bm \psi^{\dag}_{-\textbf{k} \downarrow} 
        \equiv \left( \begin{array}{c}
            \psi^{\dag}_{-\textbf{k} {yz} \downarrow} \\
            \psi^{\dag}_{-\textbf{k} {zx} \downarrow} \\
            \psi^{\dag}_{-\textbf{k} {xy} \downarrow}
        \end{array} \right).
\end{align}
The non-interacting Hamiltonian \eref{eq:nambu_non-int_ham} is now a $6 \times 6$ matrix 
in the basis of Wannier orbitals
with
\begin{equation}
    \label{eq:non-int_ham_sro}
    \hat{\mathcal{H}}_{0\textbf{k}} = \left( \begin{array}{cc} \hat{\epsilon}_{\textbf{k}} & \\ & -\hat{\epsilon}_{-\textbf{k}}^T \end{array} \right)
\end{equation}
and the number operator \eref{eq:nambu_number_op} is expressed as
\begin{equation}
    N = \sum_{\textbf{k}} \hat{\Psi}_{\textbf{k}}^\dag \left(\hat{\tau_3} \otimes \hat{\mathbb 1}_3 \right) \hat{\Psi}_{\textbf{k}}.
\end{equation}

Similarly, all objects expressed in the Nambu basis such as the source terms matrix $\hat{\phi}$, lattice Green's function $\hat{G}_{\textbf{k}}$ and Weiss field $\hat{\mathcal{G}}_0$ are now all represented by $6 \times 6$ matrices.
The superconducting components correspond to the off-diagonal $3 \times 3$ blocks.
See for example the pairing fields \eref{eq:phi_pp_ph}, for which only $\phi^{pp}$ and $\phi^{\bar{pp}}$ are non-zero and they are equal because of the relation of \eref{eq:F_Fbar}.
Similarly, the dynamical pairing responses can be represented as $3 \times 3$ matrices in spin-orbital space.
In imaginary-time space, the external pairing field $\hat{\phi}(\tau)$ is expanded in a truncated Legendre basis of rank $N$.
Denoting a spin-orbital by $\mu_i$, a general pairing field as in \eref{eq:general_phi} is now a $3 \times 3$ matrix where a component $\mu_1,\mu_2$ can be expanded in the Legendre basis as
\begin{equation}
    \phi_{\mu_1\mu_2}(\tau) = \sum_{\alpha} \phi_{\alpha;\mu_1\mu_2}  P_\alpha(\tau).
\end{equation}
Thus, the total basis for the fields in spin-orbital-Legendre, for fields with coefficients $\phi_{\alpha;\mu_1\mu_2}$,
has $N \times 3 \times 3$ elements.

The generalization for the dynamical pairing response in the linear regime \eref{eq:expansion_F} is that the $\mu_3\mu_4$ component is expressed as
\begin{align}
    F^{\mu_3\mu_4}_{\phi}(\tau) & = \sum_{\alpha\mu_1\mu_2} \phi_{\alpha;\mu_1\mu_2} \chi^{\mu_3\mu_4}_{\alpha;\mu_1\mu_2}(\tau) + \mathcal{O}(\phi^2)
\end{align}
where $\chi_{\alpha;\mu_1\mu_2}^{\mu_3\mu_4}$ is the response in the $\mu_3\mu_4$ component
obtained by considering a pairing source field $P_\alpha$ on the $\mu_1\mu_2$ spin-orbital component.
Expanded in the Legendre basis, it reads
\begin{align}
    \label{eq:gorkov_components_sro}
    \chi^{\mu_3\mu_4}_{\alpha;\mu_1\mu_2}(\tau)
        & = \sum_\beta \ \chi^{\beta;\mu_3\mu_4}_{\alpha;\mu_1\mu_2}P_\beta(\tau). 
\end{align}
Thus, the multi-spin-orbital dynamical pairing susceptibility is expressed as a matrix with components $\chi_{s}^{s'}$ where $s\equiv (\alpha, \mu_1, \mu_2)$ and $s' \equiv (\beta, \mu_3, \mu_4)$.
We will diagonalize it to find the eigenvalues and eigenvectors.

In spin-orbital space, instead of computing the response to all $3 \times 3$ components, we can use the underlying symmetries of the dynamical pairing response (proofs are provided in \sref{sec:sym_F}):
\begin{enumerate}
    \item If the external field is even-frequency (odd-frequency), the resulting dynamical pairing response will be purely even-frequency (odd-frequency).
    This is only true because we neglected spin-orbit coupling~\cite{gingras_superconductivity_2022}.
    \item The response to an inter-spin-orbital field in the $\mu_1\mu_2$ component is related to the response to the $\mu_2\mu_1$ component by
    \begin{equation}
        \chi^{\beta;\mu_3\mu_4}_{\alpha;\mu_1\mu_2} = \chi^{\beta;\mu_4\mu_3}_{\alpha;\mu_2\mu_1}.
    \end{equation}
    \item The symmetry between the $zx$ and $yz$ orbitals in the local Green's function due to the tetragonal nature of SRO leads to
    \begin{equation}
        \chi^{\beta;\bar{\mu}_3\bar{\mu}_4}_{\alpha;\bar{\mu}_1\bar{\mu}_2} = \chi^{\beta;\mu_3\mu_4}_{\alpha;\mu_1\mu_2}
    \end{equation}
    where $\bar{zx}, \bar{yz}, \bar{xy} \equiv yz, zx, xy$.
\end{enumerate}
The first symmetry can be used to reduce quantum Monte Carlo noise in the data, while the second and third reduce the external pairing field matrix from $9$ to $4$ independent components.

\begin{figure}[t]
    \centering
    \includegraphics[width=\linewidth]{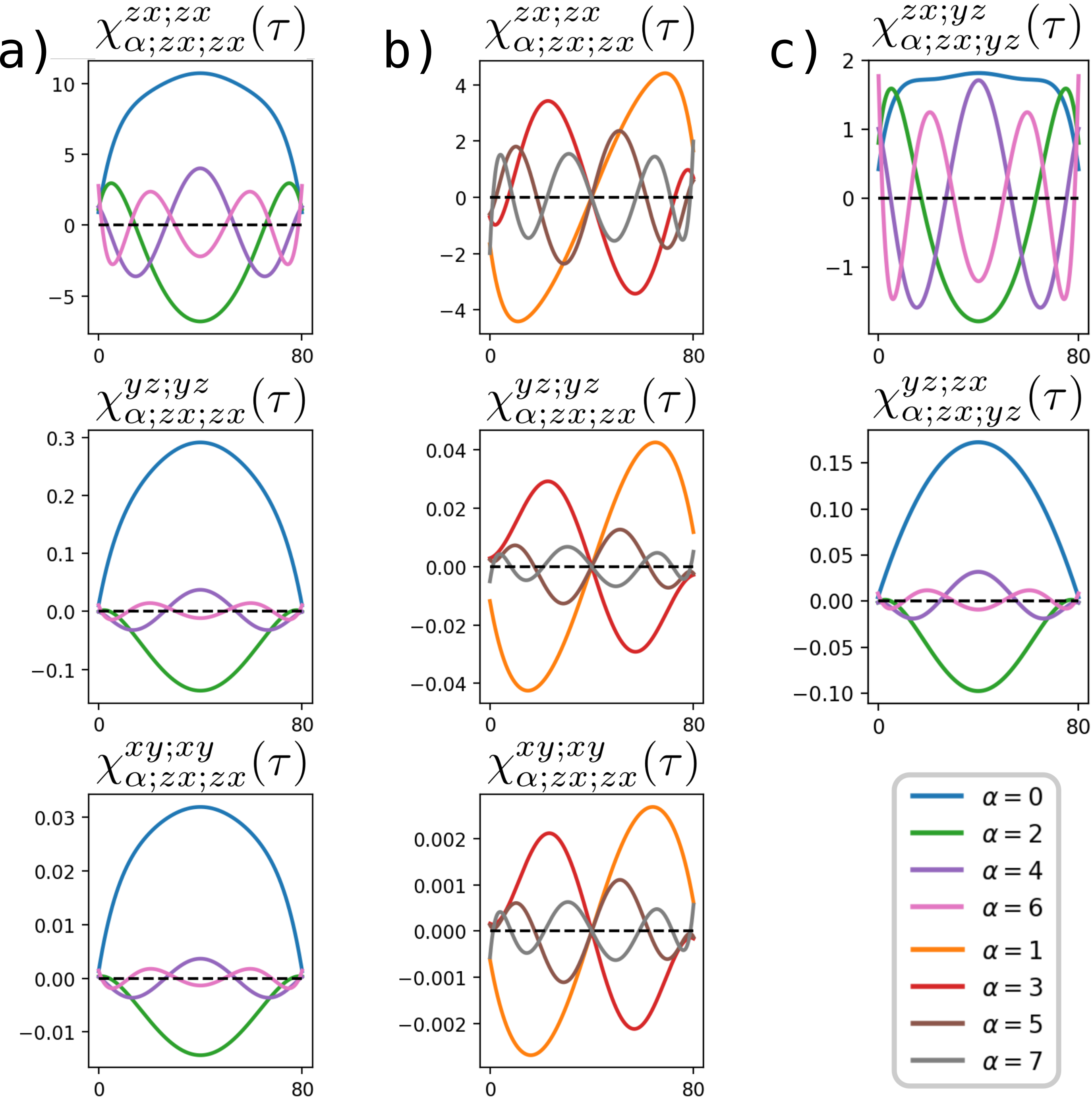}
    \caption{
    \label{fig:sro_example_F}
    Examples of dynamical pairing responses to external pairing fields in strontium ruthenate.
    Here, $\beta=80$ and $\phi_{\alpha;\mu_1\mu_2}=0.0002 \ \forall \alpha, \mu_1, \mu_2$.
    In (a) and (b), the external pairing fields are odd-frequency and even-frequency respectively, on the $zx;zx$ intra-spin-orbital component.
    The pair hopping term in the interaction induces a response in all intra-spin-orbital components.
    In (c), the external pairing fields are even-frequency on the $zx;yz$ inter-spin-orbital component.
    The spin-flip term in the interaction induces a response in the inter-spin-orbital component related by exchanging the orbital labels, here $yz;zx$.
    The components not illustrated here are smaller than $10^{-7}$
    for these fields.}
\end{figure}

Figure~\ref{fig:sro_example_F} presents some examples of dynamical pairing responses to fields.
In (a), the external fields are intra-orbital only in the $zx;zx$ sector and odd-frequency.
It is the same for (b) but with even-frequency external fields. 
According to the first symmetry detailed above, the dynamical responses share the same frequency-parity as their respective external fields.
Moreover, although these fields explicitly generate pairs only in the $zx;zx$ component, non-vanishing responses are present in all three intra-orbital components.
The explanation is that the $zx;zx$ pairs generated by the fields can tunnel to another intra-orbital component through the pair hopping term of the interaction.
This mechanism leads to pairing of smaller amplitude in these other intra-orbital components.

In \fref{fig:sro_example_F}~(c), the external fields are odd-frequency, in the inter-orbital ${zx;yz}$ sector.
Again, these fields generate stronger pairing in the $zx;yz$ component directly affected by the field.
However, there is a smaller $yz;zx$ pairing amplitude generated through the spin-flip term of the interaction.


Moreover, one can see from \fref{fig:sro_example_F} that the amplitude of the responses 
are decreasing with increasing Legendre index, an indication that the dominant responses 
are associated with the polynomials of lower index.
In other words, this indicates that eigenvectors and eigenvalues should converge quickly 
with the Legendre basis size $N$.
The convergence with $N$ is presented in \fref{fig:sro_conv_N} and discussed in \aref{sec:additional_sro}.

\renewcommand{\arraystretch}{1.5}
\begin{figure*}
  \begin{minipage}[b]{.50\linewidth}
    \centering
    \includegraphics[width=.9\linewidth]{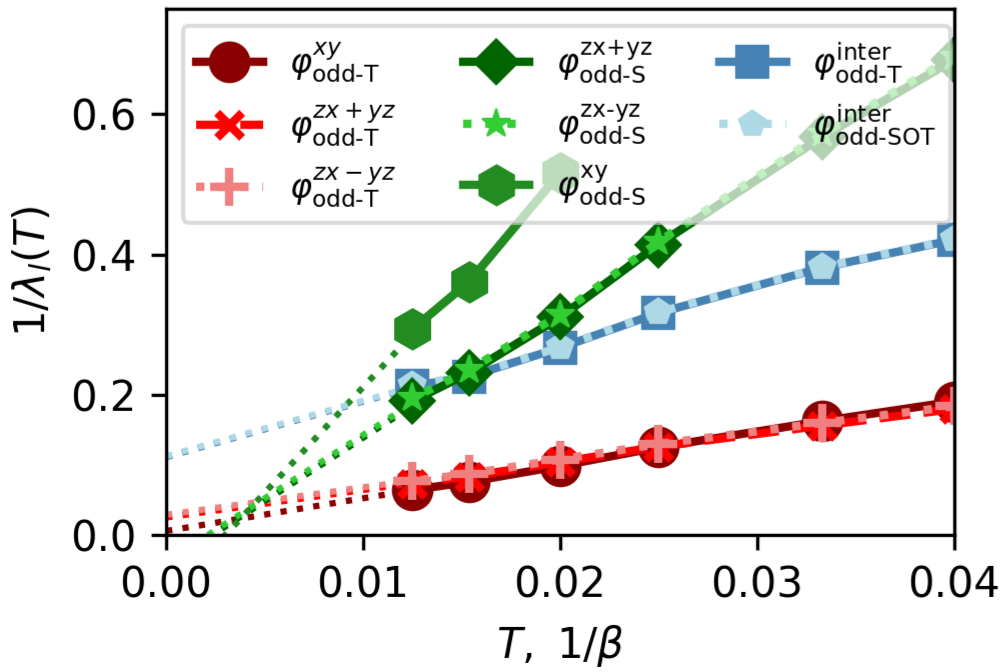}
    \par\vspace{0pt}
  \end{minipage}\hfill
  \begin{minipage}[b]{.45\linewidth}
    \centering
    \begin{tabular}{ ccc }
        \hline \hline
        \quad \quad Name \quad \quad & \quad \quad Orbital content \quad \quad & \quad \quad $SPOT$ \quad \quad \\
        \hline
        $\varphi^{xy}_{\text{odd-}T}$ & $|xy;xy\rangle$ & $^+S^+P^+O^-T$ \\
        $\varphi^{zx\pm yz}_{\text{odd-}T}$ & $|zx;zx\rangle \pm |yz;yz \rangle $ & $^+S^+P^+O^-T$ \\
        $\varphi^{xy}_{\text{odd-}S}$ & $|xy;xy\rangle$ & $^-S^+P^+O^+T$ \\
        $\varphi^{zx\pm yz}_{\text{odd-}S}$ & $|zx;zx\rangle \pm |yz;yz \rangle $ & $^-S^+P^+O^+T$ \\
        $\varphi^{\text{inter}}_{\text{odd-}T}$ & $|xy;zx\rangle \pm |xy;yz\rangle $ & $^+S^+P^+O^-T$ \\
        $\varphi^{\text{inter}}_{\text{odd-}SOT}$ & $|xy;zx\rangle \pm |xy;yz\rangle $ & $^-S^+P^-O^-T$ \\
        \hline \hline
    \end{tabular}
    \par\vspace{30pt}
  \end{minipage}
    \caption{
        \label{fig:sro_eigval_vs_T_min}
        Inverse of the dominant superconducting eigenvalues in the normal state of strontium ruthenate.
        The dotted lines are fits to find the temperatures at which the inverse eigenvalues diverge.
        Additional eigenvectors and details can be found in \ref{sec:additional_sro}.
        Note that odd-$T$ refers to the exchange of relative time and not time-reversal symmetry~\cite{gingras_superconductivity_2022, linder_odd-frequency_2019}.
    }
\end{figure*}


\subsection{\label{sec:sro_eigvecs}Results.}

We solve the eigenvalue problem associated with the dynamical pairing susceptibility matrix 
for multiple temperatures in the normal state.
Each eigenvector is characterized and labeled by
the orbitals that dominantly host their Cooper pairs and by their character under the $SPOT$ symmetry operations. 
Because we work with spinless particles, the spin parts of the Cooper pairs is deduced from the $SPOT$ condition.
We plot in Fig.~\ref{fig:sro_eigval_vs_T_min} the $T$-dependence of a few dominant eigenvalues $\lambda_l$.  
We find that each $1/\lambda_l$ depends linearly on $T$, so that a fit to this 
$T$-dependence allows to estimate the critical temperature at which 
$1/\lambda_l$ extrapolates to zero, signaling a potential superconducting phase transition.

%

The $SPOT$ classification used to characterize the eigenvectors in Fig.~\ref{fig:sro_eigval_vs_T_min} has been introduced in Ref.~\citenum{linder_odd-frequency_2019} and is reviewed 
in App.~\ref{sec:SPOT}. 
It works as follow: a state denoted by $^+S^+P^+O^-T$ is even under spin-symmetry $S$ (\textit{i.e.} a triplet in 
the absence of spin-orbit), even under parity $P$, even under orbital symmetry $O$ and odd under $T$ the exchange of the electronic times.
Note that we only have access to even-parity state (even in $\textbf{k} \rightarrow - \textbf{k}$), since 
single-site DMFT only gives us access to pairing functions which are spatially local (momentum-independent) 
in the orbital basis.
Moreover, as discussed in Ref.~\citenum{gingras_superconductivity_2022}, 
the local $t_{2g}$ orbitals transform as non-trivial irreducible representations (irreps) of the crystal point group.
As a result, the SCOP transforms as an irrep, which depends in orbital space on the different orbital labels and their relative phase.
The different possible orbital basis functions expressed in terms of irreps are presented in Table~\ref{tab:orb_basis}.



For a more  complete list of eigenvalues and a more detailed description of the superconducting 
states found in our analysis, the reader is referred to App.~\ref{sec:additional_sro}.

\renewcommand{\arraystretch}{1}
\begin{table}[h]
    \centering
    \caption{
    Basis functions in orbital space for superconducting order parameters expressed as irreducible representations (irreps).
    It depends on which orbitals are hosts to Cooper pairs, along with the relative phases with other orbital hosts. 
    $O$ is the orbital symmetry under orbital exchange.
    Table taken from Ref.~\citenum{gingras_superconductivity_2022}.
    }
    \begin{tabular}{ccc}
        \hline \hline
        \quad \quad Irrep \quad \quad \quad & \quad \quad Orbital basis function \quad \quad \quad & \quad \quad $O$ \quad \quad \quad \\
        \hline
        A$_{1g}$ & $|xy;xy\rangle$ & + \\
        A$_{1g}$ & $|yz;yz\rangle + |zx;zx\rangle$ & + \\
        B$_{1g}$ & $|yz;yz\rangle - |zx;zx\rangle$ & + \\
        A$_{2g}$ & $|yz;zx\rangle - |zx;yz\rangle$ & - \\
        B$_{2g}$ & $|yz;zx\rangle + |zx;yz\rangle$ & + \\
        E$_{g}$ & $|xy;yz\rangle + |yz;xy\rangle$ & + \\
                & $|xy;zx\rangle + |zx;xy\rangle$ & + \\
        E$_{g}$ & $|xy;yz\rangle - |yz;xy\rangle$ & - \\
                & $|xy;zx\rangle - |zx;xy\rangle$ & - \\
        \hline \hline
    \end{tabular}
    \label{tab:orb_basis}
\end{table}

\subsubsection{Intra-orbital odd-frequency states.}

\begin{figure}[b]
    \centering
    \includegraphics[width=0.9\linewidth]{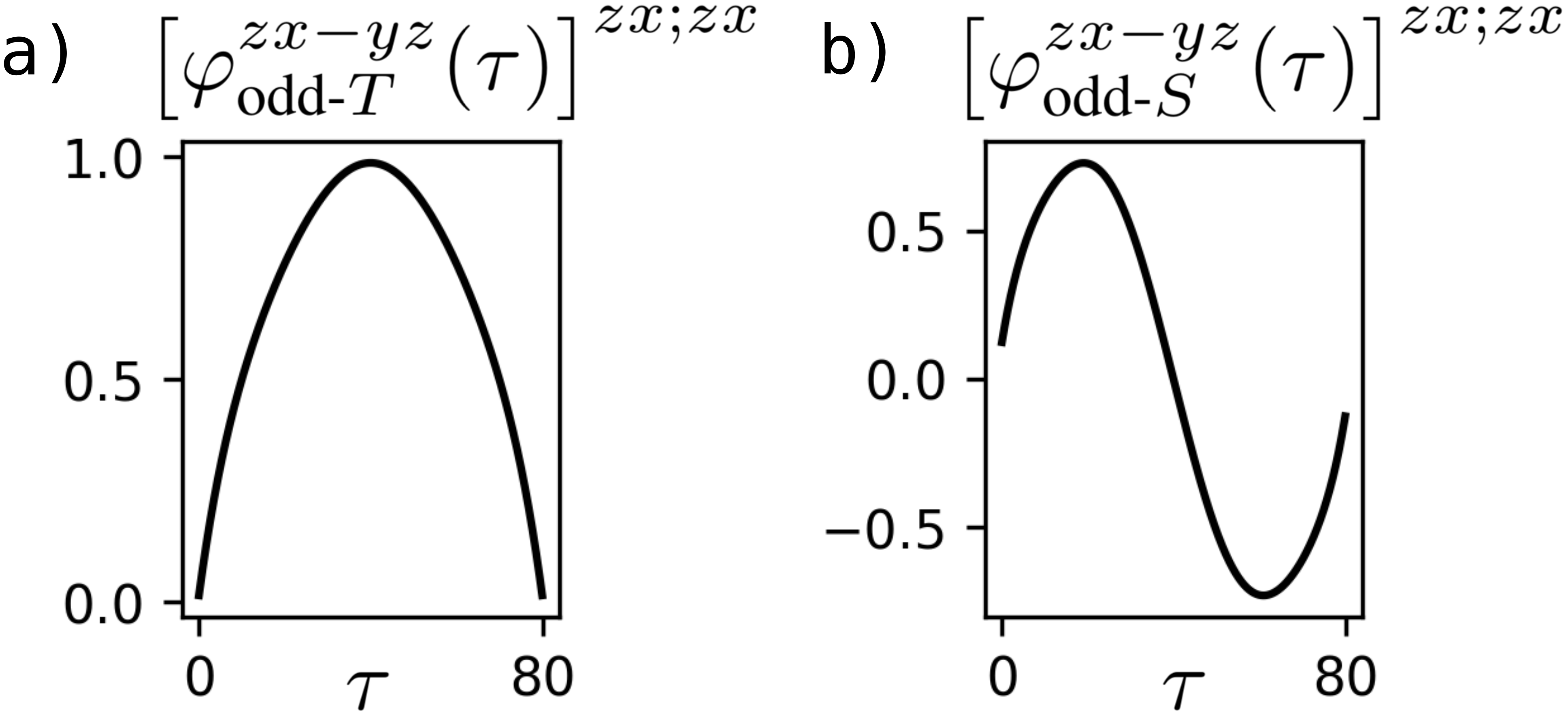}
    \caption{Imaginary-time-dependence of one of the largest components ($zx;zx$) of  the a) $\varphi^{zx\pm yz}_{\text{odd-}T}(\tau)$ and b) $\varphi^{zx\pm yz}_{\text{odd-}S}(\tau)$ states.
            These pairings take advantage of the strong retardation of the pairing interaction to avoid even-time pairing and instead utilize off-time pairing.
            Other components and different temperatures of all the dominant states are present in App.~\ref{sec:additional_sro}.}
    \label{fig:sro_f_tau_ex}
\end{figure}

The eigenvectors with by far the largest eigenvalues in the temperature range studied are $\varphi_{\text{odd-T}}^{xy}$ and $\varphi_{\text{odd-T}}^{zx\pm yz}$, labelled in red in \fref{fig:sro_eigval_vs_T_min}.
These eigenvectors are dominant because they are constrained by symmetry to have vanishing equal-time local pairing, 
hence avoiding the energy cost of the on-site local repulsion.
Instead, they form local pairs at different imaginary-times, an idea proposed by Berezinskii in the context of Helium 3~\cite{berezinskii_new_1974}. 
An example of the imaginary-time dependence of such a pairing state is displayed in Fig.~\ref{fig:sro_f_tau_ex}~(a). 
As will now be explained, some of these intra-orbital odd-frequency states correspond to the A$_{2g}^-$ state found to be a leading eigenvector in Ref.~\citenum{gingras_superconductivity_2022}.
It was argued there that it could explain the conflicting interpretations between muon spin-relaxation~\cite{grinenko_split_2021} and specific heat~\cite{li_high-sensitivity_2021}.
Again in this work, this state is found to be dominant in the range of temperatures studied.

The detailed structures of these states in orbital-space and imaginary-time along with their evolution with inverse temperature are presented in \aref{sec:additional_sro}.
These states are intra-orbital, meaning that the Cooper pairs are formed by electrons on the same orbital.
As a result, they have to be even in the exchange of orbital labels ($^+O$).
They are also even under parity ($^+P$), 
 like all local states,  and they are found to be odd-frequency ($^-T$). 
Thus, the $SPOT$ classification detailed in \sref{sec:SPOT} imposes 
that they must be spin-triplet $^+S$ and they are classified by $^+S^+P^+O^-T$.

We now discuss the irreps of the normal state's point group 
(here $D_{4h}$) according to which these states transform. 
The irrep of a given state corresponds to the product of its momentum, spin and orbital irreps.
All the irreps considered in this work being local, they all transform as the A$_{1g}$ irrep in momentum-space.
In spin-space, a spin-triplet is characterized by a $\textbf{d}$-vector~\cite{RevModPhys.75.657}.
If it point along the $z$-axis, it transforms as A$_{2g}$, and if it points in the plane, it is degenerate to another in-plane vector and both transform as E$_g$ ~\cite{gingras_superconductivity_2022}.
In the present work, we cannot distinguish between the A$_{2g}$ and E$_g$ triplets because 
they are degenerate in the absence of spin-orbit coupling.
In orbital-space, there are three possible intra-orbital components, presented as the three first basis functions of Table~\ref{tab:orb_basis}.
The intra-orbital eigenvectors of the dynamical pairing susceptibility 
form linear combinations of these three components, as long as the global irrep is the same.

Let us now discuss the orbital-space irreps.
For $\varphi_{\text{odd-T}}^{xy}$, the Cooper pairs are mostly formed by electrons on the $xy$ orbital which transforms as A$_{1g}$ in orbital-space.
The pair hoping term discussed in \sref{sec:sro_multi-orb_model} leads to small but non-vanishing intra-orbital pairing on the $yz$ and $zx$ orbitals.
To satisfy the irrep selection rule, this pairing also transforms as A$_{1g}$ in orbital-space, made possible by the $|yz;yz \rangle + |zx;zx\rangle$ basis function.
For $\varphi_{\text{odd-T}}^{zx+yz}$, the Cooper pairs are mostly formed on the $zx$ and $yz$ orbitals, in the A$_{1g}$ channel represented by the $|yz;yz \rangle + |zx;zx\rangle$ basis function.
Now pair hoping generates small non-vanishing $|xy;xy \rangle$ pairing.
Finally, $\varphi_{\text{odd-T}}^{zx-yz}$ corresponds to the third possible intra-orbital linear combination, that is the $|yz;yz \rangle - |zx;zx \rangle$ basis function.
This function transforms as B$_{1g}$ in orbital-space~\cite{gingras_superconductivity_2022} and there are no B$_{1g}$ orbital basis function involving the $xy;xy$ component.

We are now in position to discuss the global irreps.
For $\varphi_{\text{odd-T}}^{xy}$ and $\varphi_{\text{odd-T}}^{zx+yz}$, they transform as A$_{1g}$ in momentum-space, as A$_{1g}$ in orbital-space and as either A$_{2g}$ or E$_g$ in spin-space.
Thus, they globally transform as A$_{2g}$ and E$_g$.
The A$_{2g}$ solution corresponds to the one proposed in \cite{gingras_superconductivity_2022}.
As for $\varphi_{\text{odd-T}}^{zx-yz}$, it transforms as A$_{1g}$ in momentum-space, as B$_{1g}$ in orbital-space and as either A$_{2g}$ or E$_g$ in spin-space.
Thus it globally transforms as either B$_{2g}$ or E$_g$.

\begin{figure}[h]
    \centering
    \includegraphics[width=\linewidth]{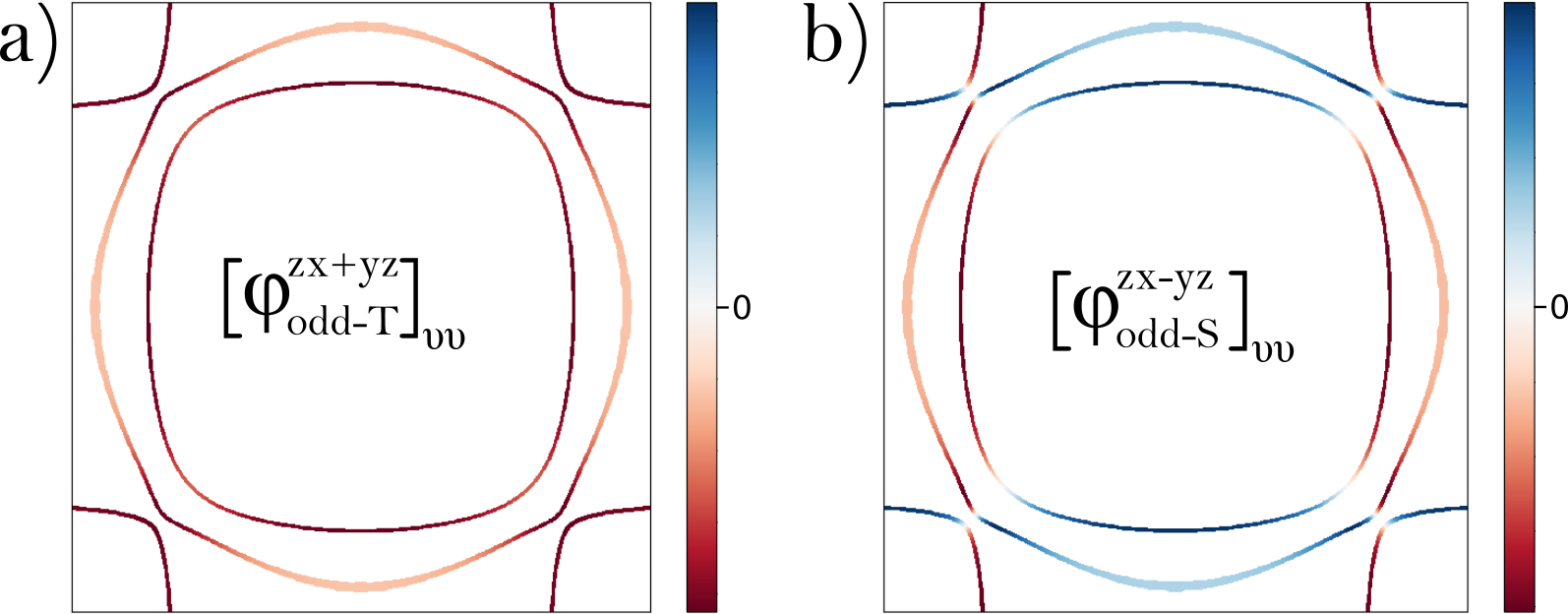}
    \caption{Projection on the Fermi surface of the gap functions (a) $\varphi_{\text{odd-}T}^{zx+yz}$ and (b) $\varphi_{\text{odd-}S}^{zx-yz}$ rotated to the band basis, for one flavor of pseudospin.}
    \label{fig:sro_gap_bands}
\end{figure}

In order to better interpret these states, we can rotate them from the orbital to the band basis of the normal state.
This operation is performed using projector matrices $P_{\textbf{k}}$ computed at each $\textbf{k}$-point, which correspond to the eigenvectors of the non-interacting Hamiltonian $\mathcal{H}_{0\textbf{k}}$ in the orbital basis.
The rows have band indices while the columns have orbital indices.
These eigenvectors allow to diagonalize the Hamiltonian into the band basis as $P^\dagger_{\textbf{k}} \mathcal{H}_{\textbf{k}0} P_{\textbf{k}}$.
They are also useful to rotate the SCOPs $\Delta$ from the orbital basis to the band basis by employing
\begin{equation}
    \label{eq:gap_band}
    \Delta_{\textbf{k}\nu\nu'} = \sum_{mm'} P_{\textbf{k}\nu m} P_{\textbf{k} \nu' m'} \Delta_{mm'}.
\end{equation}
We use the non-interacting Hamiltonian Eq.~(\ref{eq:non-int_ham_sro}) to which we manually add the spin-orbit term as described in Ref.~\citenum{tamai_high-resolution_2019}, with $\lambda = 0.2$~eV.
With the projectors obtained from this Hamiltonian, we rotate the SCOP $\varphi_{\text{odd-}T}^{zx+yz}$ to the band basis and project it on the Fermi surface, resulting in Fig.~\ref{fig:sro_gap_bands}~(a).
One can clearly observe that this gap transforms as A$_{1g}$ in momentum space.
This is expected, because this SCOP transforms as A$_{1g}$ in both momentum and orbital spaces.
Since the Hamiltonian with spin-orbit coupling is pseudospin-block-diagonal for $k_z=0$~\cite{gingras_superconductivity_2022}, we show the SCOP only for one pseudospin.
This gap being purely spin-triplet, the components for the other pseudospin are equal, which correspond to a symmetry breaking in spin-space~\cite{kaba_group-theoretical_2019}
In particular, this is a manifestation of the A$_{2g}$ irrep this state transforms as.

\subsubsection{Intra-orbital even-frequency states.}

Another set of eigenvectors found to be dominant are the $\varphi_{\text{odd-S}}^{xy}$ and $\varphi_{\text{odd-S}}^{zx\pm yz}$ states, labelled in green in \fref{fig:sro_eigval_vs_T_min}.
Although they never have larger eigenvalues than the red states, they have the fastest increase with decreasing temperature, and they end up with the largest expected transition temperatures.
These intra-orbital ($^+O$) even-parity ($^+P$) states are even-frequency ($^+T$). 
Again by the Pauli principle, they have to be spin-singlet ($^-S$) and overall transform as $^-S^+P^+O^+T$.
Their momentum, orbital and inverse temperature structures are presented in App.~\ref{sec:additional_sro}.

In terms of irreps, again these local states transform as A$_{1g}$ in momentum-space, and because they are singlets, they transform as A$_{1g}$ in spin-space.
In orbital-space, just as the intra-orbital odd-frequency states, they are composed of the three possible linear combinations of intra-orbital basis functions: $\varphi_{\text{odd-S}}^{xy}$ is dominated by A$_{1g}$ pairs on the $xy$ orbital, $\varphi_{\text{odd-S}}^{zx+ yz}$ by A$_{1g}$ pairs on the $zx$ and $yz$ orbitals and $\varphi_{\text{odd-S}}^{zx- yz}$ by B$_{1g}$ pairs on the $zx$ and $yz$ orbitals.
As a result, globally, $\varphi_{\text{odd-S}}^{xy}$ and $\varphi_{\text{odd-S}}^{zx+ yz}$ transform as A$_{1g}$ while $\varphi_{\text{odd-S}}^{zx- yz}$ transforms as B$_{1g}$.

To understand how the B$_{1g}$ irrep of the orbitals manifests on the SCOP in the band basis, we rotate the state $\varphi_{\text{odd-}S}^{zx-yz}$ to the band basis using Eq.~\ref{eq:gap_band} and project it on the Fermi surface, resulting in Fig.~\ref{fig:sro_gap_bands}~(b).
We observe that after converting the orbital indices, which transform as a non-trivial irrep, to band indices, which transform as the trivial irrep, the non-triviality is acquired by the momentum space due to the projectors $P_{\textbf{k}}$.
Note that this effect cannot happen for the $xy;xy$ component since its only orbital basis function transforms trivially.
Thus to transform as B$_{1g}$ as was found in other works~\cite{gingras_superconducting_2019, Acharya2019, gingras_superconductivity_2022}, the $xy;xy$ component requires a non-trivial momentum structure factor which is not accessible within the DMFT framework.
Nonetheless, we present a purely local solution in orbital space which actually behaves as a $d$-wave state once upfolded to the band basis.
This upfold-induced momentum dependence has also been noticed about the self-energy obtained from DMFT, which accurately reproduces the Fermi surface in the normal state~\cite{tamai_high-resolution_2019}).
Now in spin-space, this gap is purely spin-singlet and thus the components for the other pseudospin have a minus sign, which satisfies the symmetries of the spin projected on the $z$-axis and transforms as A$_{1g}$.

It is surprising to find these solutions dominating because one would generally not expect a strongly repulsive systems of electrons to favor local, intra-orbital states with spin-singlet pairing.
However, as can be seen from the imaginary-time structure of these eigenvectors, for example in Fig.~\ref{fig:sro_f_tau_ex}~(b), the equal-time pairing $\varphi(\tau=0)$ is minuscule.
This means that, similarly to the odd-frequency states, these states take advantage of the strong retardation 
to avoid equal-time local pairing and instead generate retarded pairing, which cannot be captured by static mean-field methods.
As a result, such local states would often be overlooked, 
but the general method presented in this work allows to capture them.

One possibility is that these states correspond to downfolded versions of momentum-dependent solutions such as the leading $d$-wave B$_{1g}$ candidate that explains many thermodynamic measurements~\cite{hassinger_vertical_2017, li_high-sensitivity_2021, Li2022} and that was reported in various numerical works such as in Refs~\citenum{gingras_superconductivity_2022, hauck2023competition}.
This possibility is supported by the fact that one of the states already transforms as $d$-wave in the band basis.
This cannot however be guarantee in this work, because of the other almost degenerate states transform as $s$-wave.
In order to investigate this question, we plan on including non-local pairing solutions by combining the method presented in this manuscript to cluster generalizations of DMFT, to diagrammatic extensions of DMFT or to other non-local methods. 
In general, more work is needed to better understand the conditions favoring the emergence of these off-time pairing states, along with providing possible experimental signatures of their retardation effects.

\subsubsection{Inter-orbital odd-frequency states.}

The last eigenvectors of interest are the states $\varphi_{\text{odd-}T}^{\text{inter}}$ and $\varphi_{\text{odd-}SOT}^{\text{inter}}$, labelled in blue in \fref{fig:sro_eigval_vs_T_min}.
Each of these states includes two degenerate states that transform as the E$_g$ irreducible representation with $|xy;yz\rangle$ and $|xy;zx\rangle$ in and out of phase (labelled as $\pm$), shown in Table~\ref{tab:orb_basis}.
Moreover, we find that the eigenvalues associated to them are also numerically degenerate.
This is surprising since, although these states are odd-frequency ($^-T$) with the same inter-orbital hosts, the states $\varphi_{\text{odd-}T}^{\text{inter}}$ are spin-triplet ($^+T$) even-orbital ($^+O$) while the states $\varphi_{\text{odd-}SOT}^{\text{inter}}$ are spin-singlet ($^-S$) odd-orbital ($^-O$).

The states $\varphi_{\text{odd-}SOT}^{\text{inter}}$ ($^-S^+P^-O^-T$) correspond to the leading states found in Ref.~\citenum{PhysRevB.105.155101} at temperatures between $0.025$ and $0.04$ in our units.
Since both methods start from similar non-interacting Hamiltonians and, in principle, construct the pairing glue using DMFT, we would expect to find similar results.
However, we do not find these solutions to have the largest eigenvalues and their predicted critical temperature is far in the negatives.
The main conceptual difference comes in Ref.~\citenum{PhysRevB.105.155101} from an additional approximation performed on the vertex when constructing the ladder functions.
Our results suggest that this approximation leads to drastic changes in the leading eigenvectors and should be carefully investigated.


\section{\label{sec:conclu}Conclusion}

In this manuscript, we presented a `two-time linear response' method to probe pairing states with 
frequency dependence and retardation. 
The method generalizes that of Ref.~\citenum{Georges_1993} and in particular introduces an expansion of the 
two-time pairing susceptibility on a basis of Legendre polynomials. 
By inspecting the eigenvalues and eigenvectors of the susceptibility matrix in this basis, the 
potential superconducting instabilities signalled by a diverging eigenvalue can be analyzed 
and the superconducting critical temperature estimated.
This method enables to capture the frequency dependence of the superconducting order parameter and gives access to a complete description of pairing states that take advantage of retardation, \textit{i.e.} pairing states in which the two electrons follow each other 
with a time lag. 
This is especially important in strongly correlated systems for which equal-time on-site pairing is suppressed by 
local repulsive interactions, and it allows in particular to capture odd-frequency states which are impossible to study 
with static pairing fields.

We employed this method within the single-site dynamical mean-field theory framework, hence restricting our analysis to 
pairing states that are spatially local (momentum independent) when expressed in the basis of local orbitals or Wannier functions. 
First, we benchmarked this method by reproducing known results on the attractive single orbital Hubbard model, emphasizing the retardation of the order parameter down to $T_c$ at strong coupling.

Then, we applied the method to study the superconducting order parameter of strontium ruthenate in the absence of spin-orbit coupling, a system where local odd-frequency solutions have been proposed as potential candidates~\cite{komendova_odd-frequency_2017, gingras_superconducting_2019, gingras_superconductivity_2022}.
Our results corroborate that the A$_{2g}^-$ odd-frequency triplet state reported in Refs~\citenum{gingras_superconducting_2019, gingras_superconductivity_2022} is dominant at all temperatures studies.
They also suggest that the odd-frequency state reported in Ref.~\citenum{PhysRevB.105.155101} is not a competing candidate, highlighting that the alternative Eliashberg equations approach is exceedingly sensitive to the approximation made when 
constructing the particle-particle vertex function. 

Furthermore, we find that the states with highest predicted transition temperatures are local even-frequency singlet states 
that also involve strongly retarded unequal-time pairing.
We speculate that these unequal-time singlet pairing states could be downfolded representations of the momentum-dependent $d$-wave B$_{1g}$ state.
This is supported by the fact that one of these states actually transforms as $d$-wave due to the upfolding of the superconducting order parameters from the orbital basis to the band basis.
Such a state is presently the leading contending symmetry for superconducting strontium ruthenate because it can explain all thermodynamic measurements. 

Future works should focus on implementing this method while allowing for momentum-dependence. 
This can be achieved either through cluster generalizations or employing diagrammatic (vertex-based) expansions 
of dynamical mean-field theory.
This method should also be used to better understand the emergence of retarded pairing states, for which experimental signatures should be further studied.

\begin{acknowledgments}

We are grateful for discussions with Sophie D. Beck, Philipp Hansmann, Jason Kaye, Hugo U. R. Strand and Nils Wentzell. 
The Flatiron Institute is a division of the Simons Foundation.

\end{acknowledgments}

\appendix

\section{\label{sec:SPOT}Odd-frequency and $SPOT$ representation.}
The anomalous Green's function in imaginary time is
\begin{equation}
    F_{l_1l_2}^{\sigma_1\sigma_2}(\bm r_1, \bm r_2; \tau_1, \tau_2) \equiv \langle \mathcal{T}_{\tau} \psi_{l_1}^{\sigma_1}(\bm r_1; \tau_1) \psi_{l_2}^{\sigma_2}(\bm r_2; \tau_2) \rangle
\end{equation}
where $l_i$, $\sigma_i$, $\bm r_i$ and $\tau_i$ are respectively the orbital, spin, position and imaginary time indices of the electrons $i$ destroyed by $\psi^{\sigma_i}_{l_i}(\bm r_i; \tau_i)$, and $\mathcal{T}_{\tau}$ is the time ordering operator.
Note that in the main text, we combine the orbital $l$ and spin $\sigma$ indices into a single spin-orbital $\mu \equiv (\sigma, l)$ index.
Here we separate them for clarity.
Using the anticommutation property of the destruction operators in a time-ordered product, we have
\begin{equation}
    \label{eq:Pauli}
    F_{l_1l_2}^{\sigma_1\sigma_2}(\bm r_1, \bm r_2; \tau_1, \tau_2) = - F_{l_2l_1}^{\sigma_2\sigma_1}(\bm r_2, \bm r_1; \tau_2, \tau_1).
\end{equation}
Instead of using the individual positions and times, we can define the center of mass and relative positions and times
\begin{equation}
    \bm R \equiv \frac{\bm r_1 + \bm r_2}{2}, \ \bm r \equiv \frac{\bm r_1 - \bm r_2}{2} ; \ T = \tau_1 + \tau_2, \ \tau \equiv \tau_1 - \tau_2.
\end{equation}
We restrict our analysis to $\bm R=T=0$.
With these quantum number, the notation becomes
\begin{equation}
    F^{\sigma_1\sigma_2}_{l_1l_2}(\bm r_1,\bm r_2; \tau_1, \tau_2) \rightarrow F^{\sigma_1\sigma_2}_{l_1l_2}(\bm r; \tau).
\end{equation}
Applying a Matsubara transformation on the object allows to express it in Matsubara frequency, with transformation and inverse defined as
\begin{align}
    F^{\sigma_1\sigma_2}_{l_1l_2}(\bm r; i\omega_n) = \int_0^{\beta} d\tau \ e^{i\omega_n \tau} F^{\sigma_1\sigma_2}_{l_1l_2}(\bm r; \tau), \\
    F^{\sigma_1\sigma_2}_{l_1l_2}(\bm r; \tau) = \frac{1}{\beta} \sum_n e^{-i\omega_n \tau} F^{\sigma_1\sigma_2}_{l_1l_2}(\bm r; i\omega_n).
\end{align}

Using the Fourier transformation, we have
\begin{equation}
    \label{eq:pre_SPOT}
    F_{l_1l_2}^{\sigma_1\sigma_2}(\bm r; i\omega_n) = - F_{l_2l_1}^{\sigma_2\sigma_1}(-\bm r; -i\omega_n).
\end{equation}

Defining operators that exchanges every quantum number independently as
\begin{align}
    \left[\hat{S} F \right]^{\sigma_1\sigma_2}_{l_1l_2}(\bm r; i\omega_n) & \equiv F^{\sigma_2\sigma_1}_{l_1l_2}(\bm r; i\omega_n), \\
    \left[\hat{P} F \right]^{\sigma_1\sigma_2}_{l_1l_2}(\bm r; i\omega_n) & \equiv F^{\sigma_1\sigma_2}_{l_1l_2}(-\bm r; i\omega_n), \\
    \left[\hat{O} F \right]^{\sigma_1\sigma_2}_{l_1l_2}(\bm r; i\omega_n) & \equiv F^{\sigma_1\sigma_2}_{l_2l_1}(\bm r; i\omega_n) \quad \text{and} \\
    \label{eq:SPOT_oddT}
    \left[\hat{T} F \right]^{\sigma_1\sigma_2}_{l_1l_2}(\bm r; i\omega_n) & \equiv F^{\sigma_1\sigma_2}_{l_1l_2}(\bm r; -i\omega_n),
\end{align}
\eref{eq:pre_SPOT} can be expressed as the $\hat{S}\hat{P}\hat{O}\hat{T}$ condition, that is $\hat{S}\hat{P}\hat{O}\hat{T} F = - F$~\cite{linder_odd-frequency_2019, gingras_superconductivity_2022}.
Because they are involutive
if the quantum numbers are decoupled, every of these operators have eigenvalues $\pm 1$.
If the Gorkov function an eigenvector of $\hat{T}$ with eigenvalue $-1$, it is called odd-frequency.
Since we are dealing with the impurity Green's function within DMFT, we will drop the spatial index $\bm r$.

\section{\label{sec:leg}Legendre polynomials.}

In this section, we provide the definition and properties of the Legendre polynomials. The $\alpha^{th}$ Legendre polynomial is given by

\begin{equation}
    \label{eq:leg}
    \tilde{P}_\alpha(x) = \frac{1}{2^n n!} \frac{d^n}{dx^n}(x^2-1)^n
\end{equation}
and these polynomials satisfy the following orthogonality relation: 
\begin{equation}
    \int_{-1}^1 dx \ \tilde{P}_\alpha(x) \ \tilde{P}_\gamma(x) = \frac{2}{2\alpha+1} \delta_{\alpha\gamma}.
\end{equation}

In the main text, we use the Legendre basis to represent Green's function in imaginary-time.
Those are defined between $0$ and $\beta$. We also prefer the orthogonality relation to be orthonormal.
Thus, we use the following rescaled Legendre polynomials:
\begin{equation}
    P_\alpha(\tau) = \sqrt{\frac{2\alpha+1}{2}} \ \tilde{P}_{\alpha}\left(\frac{\beta}{2}(x+1)\right).
\end{equation}
The orthogonality relation becomes
\begin{equation}
    \label{eq:leg_ortho}
    \frac{2}{\beta} \int_{0}^\beta d\tau \ P_\alpha(\tau) \ P_\gamma(\tau) = \delta_{\alpha\gamma}.
\end{equation}

\begin{figure}[t]
    \centering
    \includegraphics[width=\linewidth]{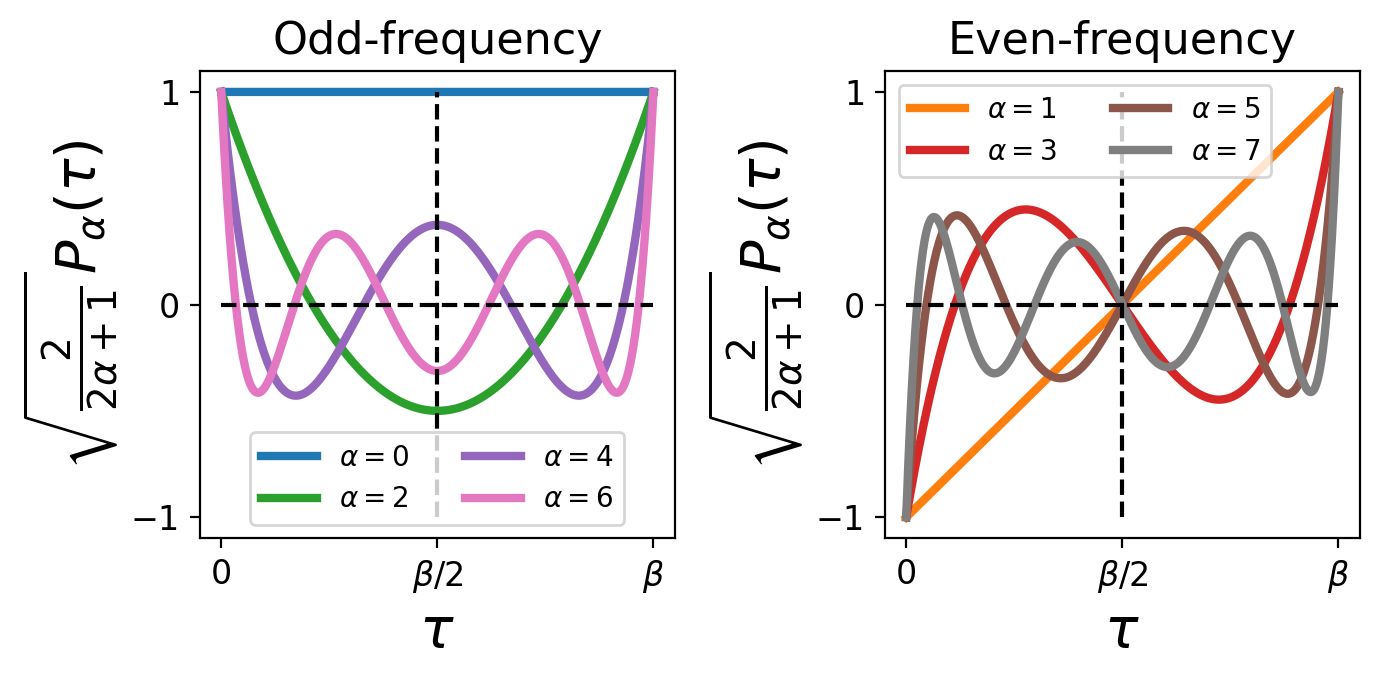}
    \caption{First eight rescaled Legendre polynomials $P(\tau)$.}
    \label{fig:leg_poly}
\end{figure}

The first eight rescaled Legendre polynomials are plotted in \fref{fig:leg_poly}.
An arbitrary, for example the Green's function, can be expanded in the rescaled Legendre basis as
\begin{equation}
    G(\tau) = \sum_\alpha G_\alpha P_\alpha(x).
\end{equation}
To calculate the coefficients $G_\alpha$, we use the orthogonality relation as
\begin{equation}
    G_\alpha = \frac{2}{\beta} \int_{0}^\beta d\tau \ G(\tau) \ P_\alpha(\tau).
\end{equation}

\section{\label{sec:sym_beta_2}Symmetry of the $F(\tau)$ around \texorpdfstring{$\beta/2$}{beta/2}.}
Let's consider the Gorkov function at $\beta/2+\tau$.
Following the notation of \ref{sec:SPOT} with spin-orbital $\mu \equiv (\sigma, l)$ and using time-translational invariance, we can write
\begin{equation}
    F^{\mu_1 \mu_2}(\beta/2 + \tau) = \langle \mathcal{T}_\tau \psi^{\mu_1}(\beta/2) \psi^{\mu_2}(-\tau) \rangle.
\end{equation}

The expectation value being defined as 
\begin{equation}
    \langle \mathcal{O} \rangle \equiv \frac{1}{Z} \text{Tr} \left[ e^{-\beta \mathcal{H}} \mathcal{O} \right],
\end{equation}
we can use the cyclicity of the trace to obtain
\begin{equation}
    F^{\mu_1 \mu_2}(\beta/2 + \tau) = \langle \mathcal{T}_{\tau} \psi^{\mu_2}(\beta/2-\tau) \psi^{\mu_1} \rangle.
\end{equation}

According to the definition of the time-ordering operator and using time-translational invariance, we have
\begin{equation}
    F^{\mu_1 \mu_2}(-\tau) = - \langle \mathcal{T}_{\tau} \psi^{\mu_2}(\tau) \psi^{\mu_1} \rangle.
\end{equation}
Thus,
\begin{equation}
    F^{\mu_1 \mu_2}(\beta/2 + \tau) = -F^{\mu_1 \mu_2}(-\beta/2 + \tau).
\end{equation}

In other words, the symmetry around $\beta/2$ depends on the even or oddness under $\hat{T}$, that is
\begin{align}
    F^{\mu_1 \mu_2}\left( \beta/2 + \tau \right) = \mp F^{\mu_1 \mu_2}\left( \beta/2 - \tau \right) \\
    \quad \quad \text{if} \ \ F^{\mu_1 \mu_2}(-\tau) = \pm F^{\mu_1 \mu_2}(\tau). \nonumber
\end{align}
It is equivalent to
\begin{equation}
    \label{eq:parity_freq_eq_beta}
    F^{\mu_1 \mu_2}\left( \tau \right) = \mp F^{\mu_1 \mu_2}\left( \beta - \tau \right).
\end{equation}

\section{\label{sec:sym_F}Symmetries of the dynamical pairing response.}

We obtain properties of the dynamical pairing response by exploiting the fact it is defined by the second functional derivative of the free-energy with respect to source fields, around the solution without sources fields, that is
\begin{equation}
    \chi^{pp}(\tau, \tau')
    \equiv \frac{\delta^2 \mathcal{F}_\phi}{\delta {\phi}^{\bar{pp}}(\tau)\delta {\phi}^{pp}(\tau')} \Big\lvert_{\phi=0}
    \hspace{-.2cm} = \frac{\delta F_\phi(\tau)}{\delta {\phi}^{pp}(\tau')} \Big\lvert_{\phi=0}.
    \label{eq:pp_suscep}
\end{equation}
We used the fact that the anomalous Green's function is also defined as a functional derivative of the free-energy.
This susceptibility is often defined with a factor $2$ because of the indistinguishability of the particles in the particle-particle channel which is not present in the particle-hole channel~\cite{bickers_self-consistent_2004, gingras_superconductivity_2022}.
In this work, we absorbed this factor to simplify the expressions.

\paragraph*{Imaginary-time.}
Presuming an external pairing field $\hat{\phi}_{e/o}$ is even/odd in frequencies, by \eref{eq:parity_freq_eq_beta} it also satisfies
\begin{equation}
    \hat{\phi}_{e/o}(\beta-\tau) = \mp \hat{\phi}_{e/o}(\tau).
\end{equation}
According to \eref{eq:pp_suscep},
\begin{align}
    \chi^{pp}_{e/o}(\beta - \tau, \tau'; \phi) & =  \frac{\partial^2 \mathcal{F}[\phi]}{\partial \hat{\phi}_{e/o}(\beta - \tau) \partial \hat{\phi}_{e/o}(\tau')} \nonumber \\
    &  = \mp \chi_{e/o}^{pp}(\tau, \tau'; \phi).
\end{align}
Thus, the anomalous Green's function given by \eref{eq:lin_resp} in the linear regime satisfies the same even/odd-frequency as the external pairing field it's responding to, that is
\begin{equation}
    F_{\phi^{e/o}} (\beta - \tau) = \mp F_{\phi^{e/o}} (\tau).
\end{equation}
This property is thus also transferred to the linear dynamical pairing susceptibility $\chi_\alpha(\tau)$ in \eref{eq:expansion_F}.

\paragraph*{Spin-orbitals.}

In the normal state, the $yz$ and $zx$ orbitals are degenerate and related by symmetry. 
We define the symmetry acting on the orbitals as $\bar{a} = xy, zx, yz$ for $a = xy, yz, zx$.
The spin-orbital components of the pairing response function, \eref{eq:gorkov_components_sro}, have to satisfy the following relation:
\begin{equation}
    \chi_{\alpha;\bar{\mu}_1\bar{\mu}_2}^{\bar{\mu}_3\bar{\mu}_4} = \chi_{\alpha;\mu_1\mu_2}^{\mu_3\mu_4}.
\end{equation}

\section{\label{sec:lin_regime}Linear regime for the Hubbard model.}

\begin{figure}[h]
    \centering
    \includegraphics[width=\linewidth]{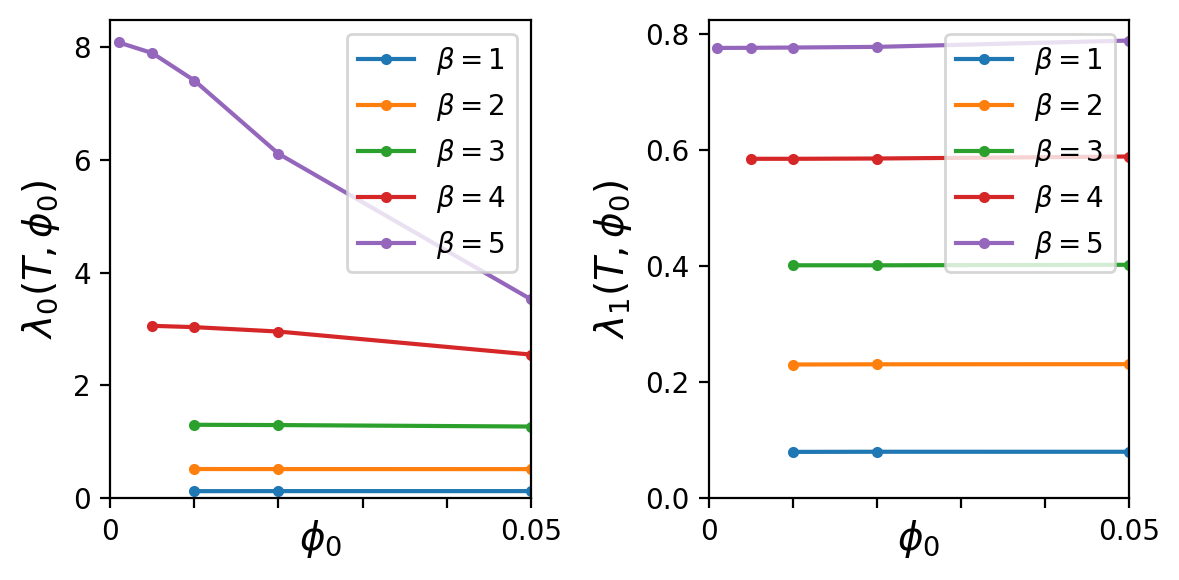}
    \caption{
    Convergence with respect to $\phi_\alpha = \phi_0 \ \forall  \ \alpha$ of the a) leading and b) subleading eigenvalues, at different inverse temperatures $\beta$. 
    At larger $\beta$, $\phi_0$ needs to be reduced to remain in the linear regime.
    }
    \label{fig:att_eigval_vs_phi0}
\end{figure}

With lower temperatures, the superconducting fluctuations become enhanced and eventually lead to a phase transition.
As a result, it is increasingly difficult to remain in the linear regime and the prefactors $\phi_\alpha$ needs to be adapted to $\beta$.
We show in \fref{fig:att_eigval_vs_phi0} the dependence of the (a) first and (b) second leading eigenvalues as a function of the prefactor $\phi_\alpha$ for different temperatures in the normal state.
Since the first eigenvalue is almost divergent at $\beta=5$, the external field need to be smaller than smaller values of $\beta$ to remain in the linear regime.
As for the second eigenvalue, it is far from diverging the remains in the linear regime even for larger amplitudes of the external field.

\begin{figure}[b]
    \centering
    \includegraphics[width=0.8\linewidth]{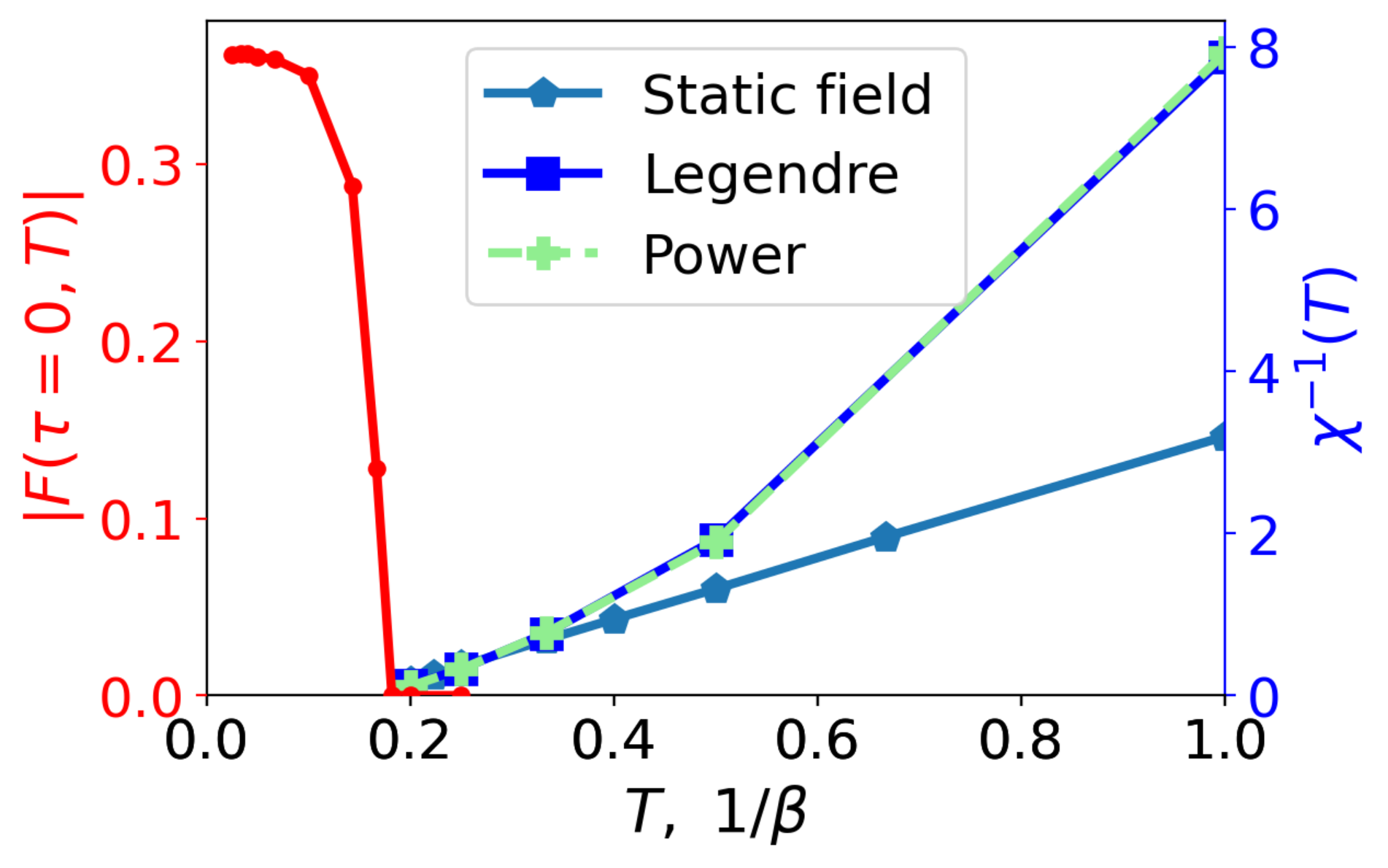}
    \caption{Comparison of the eigenvalues obtained using the power method (light green crosses) and diagonalizing the dynamical pairing susceptibility constructed in the Legendre basis (blue squares). The other components of this figure were presented in \fref{fig:att_eigval_vs_T}.}
    \label{fig:att_eigval_vs_T_with_power}
\end{figure}

\renewcommand{\arraystretch}{1.5}
\begin{figure*}
  \begin{minipage}[b]{.55\linewidth}
    \centering
    \includegraphics[width=1\linewidth]{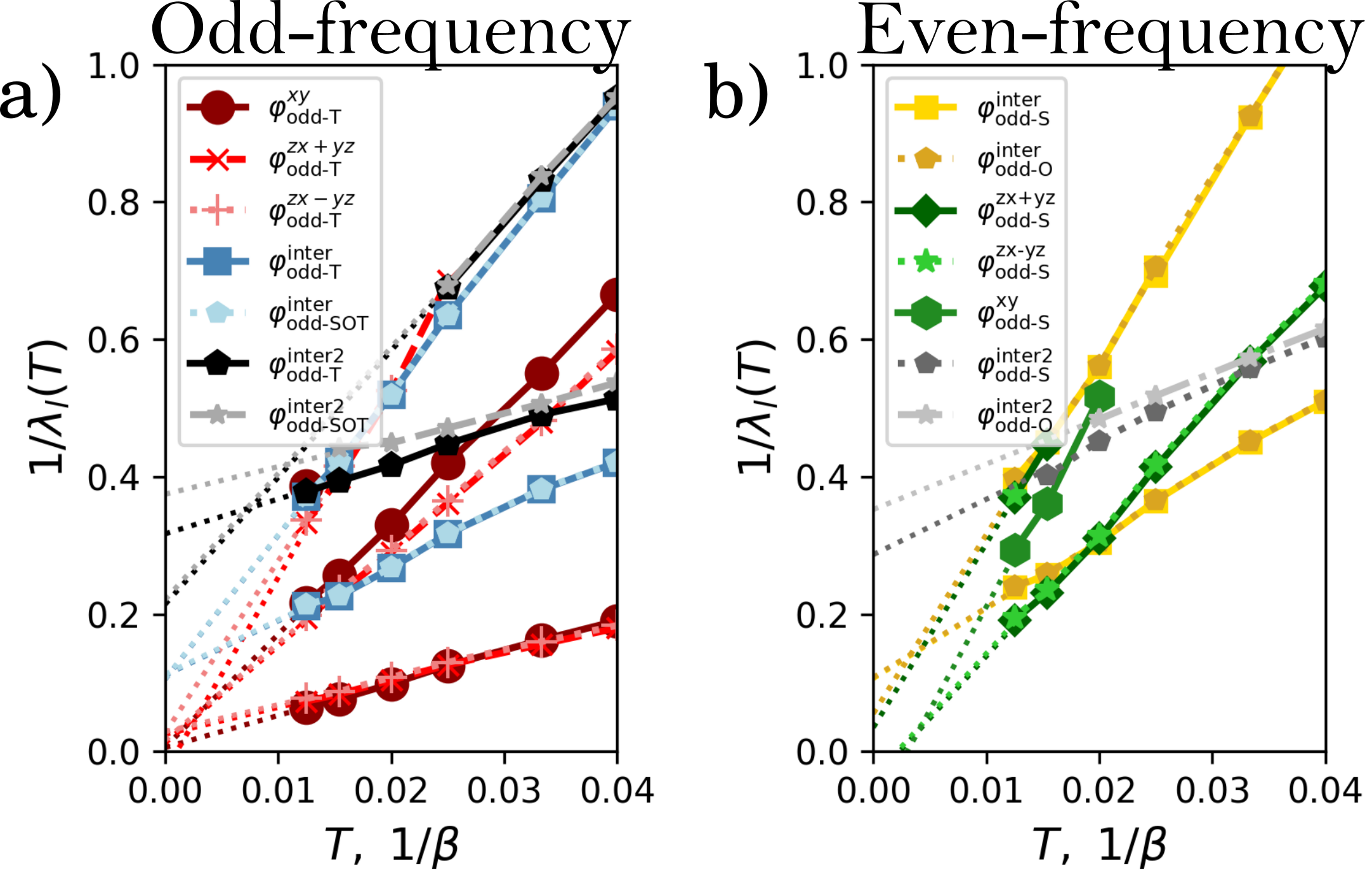}
    \par\vspace{0pt}
  \end{minipage}\hfill
  \begin{minipage}[b]{.40\linewidth}
    \centering
    \begin{tabular}{ ccc }
        \hline \hline
        \quad \quad Name \quad \quad & \quad \quad Orbital content \quad \quad & \quad \quad $SPOT$ \quad \quad \\
        \hline
        $\varphi^{\text{inter}}_{\text{odd-}S}$ & $|xy;zx\rangle \pm |xy;yz\rangle $ & $^-S^+P^+O^+T$ \\
        $\varphi^{\text{inter}}_{\text{odd-}O}$ & $|xy;zx\rangle \pm |xy;yz\rangle $ & $^+S^+P^-O^+T$ \\
        $\varphi^{\text{inter2}}_{\text{odd-}S}$ & $|zx;yz \rangle$ & $^-S^+P^+O^+T$ \\
        $\varphi^{\text{inter2}}_{\text{odd-}T}$ & $|zx;yz \rangle$ & $^+S^+P^+O^-T$ \\
        $\varphi^{\text{inter2}}_{\text{odd-}O}$ & $|zx;yz \rangle$ & $^+S^+P^-O^+T$ \\
        $\varphi^{\text{inter2}}_{\text{odd-}SOT}$ & $|zx;yz \rangle$ & $^-S^+P^-O^-T$ \\
        \hline \hline
    \end{tabular}
    \par\vspace{30pt}
  \end{minipage}
    \caption{
        \label{fig:sro_eigval_vs_T}
        Temperature dependence of the inverse of the dominant superconducting eigenvalues in the normal state of strontium ruthenate.
        Figure~(a) corresponds to the odd-frequency channel and (b) to the even-frequency channel.
        The dotted lines are fits to find the temperatures at which the inverse eigenvalues diverge.
        Figure~\ref{fig:sro_leadvec} of App.~\ref{sec:additional_sro:eigvec_beta} presents the temperature dependence and orbital structure of
        the three dominant eigenvectors in each channel.
        The dotted lines are fits to find the temperatures at which the inverse eigenvalues diverge.
        Different lines with the same color correspond to eigenvectors with the same symmetry, but orthogonal in imaginary-time.
        An example is discussed in App.~\ref{sec:additional_sro:higher-order}.
    }
\end{figure*}

\section{\label{app:power}Power method.}
Another way to obtain the leading eigenvector of the dynamical susceptibility is to use the power method.
The power method consists of starting with a random normalized field $\phi^0(\tau)$ multiplied by a small coefficient $\alpha$ to remain in the linear regime and compute the dynamical response to that field.
In the linear regime where we have \eref{eq:lin_resp}, we find
\begin{equation}
    F^0(\tau) \equiv F_{\alpha\phi^0}(\tau) = \int d\tau' \ \chi^{pp}(\tau, \tau') \alpha\phi^0(\tau') + \mathcal{O}(\alpha^2).
\end{equation}
We normalize this resulting response and then use it as the next field:
\begin{equation}
    \phi^1(\tau) = \frac{F^0(\tau)}{\int d\tau' |F^0(\tau')|} \equiv \frac{F^0(\tau)}{P^0}.
\end{equation}
Solving \eref{eq:lin_resp} with this field gives us $F^1(\tau)$.

This procedure is performed until convergence of the dynamical response, which then corresponds to the leading eigenvector $\varphi_0(\tau)$ of $\chi^{pp}$ with eigenvalue $\lambda_0$.
It works because since $\phi^0$ is random, it can by expressed as a linear combination of the eigenvectors $\varphi_l$:
\begin{equation}
    \phi^0(\tau) = \sum_l \beta_l \varphi_l(\tau) \ \ \text{where} \ \ \sum_l |\beta_l|^2 = 1
\end{equation}
and where te $\beta_l$ are the coefficient of the linear combination.
Successively applying the procedure mentioned above, one finds
\begin{align}
    \phi^N (\tau) = \frac{\alpha^{N}}{\prod_{i=0}^{N-1} P_i} \sum_l \beta_l \lambda_l^N \varphi_l(\tau).
\end{align}

Taking $N$ infity, the component with largest eigenvalue $l=0$ is the only one that remains in the non-degenerate case and since these vectors are normalized, we have
\begin{align}
    & \lim_{N\rightarrow \infty} \phi^N(\tau) = \varphi_0(\tau) \quad \text{and} \\
    & \lim_{N\rightarrow\infty} \frac{F_{\alpha\phi^N}(\tau)}{\alpha \phi^N(\tau)} = \lambda_0.
\end{align}

The results of this procedure compared with those of the static and dynamical method are shown in \fref{fig:att_eigval_vs_T_with_power}.
The light green crosses correspond to leading eigenvalues $1/\lambda_0$ obtained using the power method.
They match those obtained by diagonalizing the dynamical pairing susceptibility.
The eigenvectors, although not shown, also match.

The disadvantages of the power method are two-fold: first it only gives access to the leading eigenvalue and second it works well only when the leading eigenvalue is dominant over the others.
In a case with multiple degrees of freedom like strontium ruthenate, there are many competing states and this procedure has trouble converging.

\section{\label{sec:additional_sro}Additional results for SRO.}

The results presented in Sec.~\ref{sec:sro} of the main text represent only a small fraction of what we were able to obtain.
In this section, we present the a larger set of eigenvectors shown in Fig.~\ref{fig:sro_eigval_vs_T}, we discuss the convergence of the eigenvalues with respect to the number of Legendre polynomials $N$ in Sec.~\ref{sec:additional_sro:basis_size_conv}, we introduce the eigenvectors that are of higher-order in imaginary-time in Sec.~\ref{sec:additional_sro:higher-order} and we give more details about the temperature dependence and orbital structure of the eigenvectors discussed in the main text in Sec.~\ref{sec:additional_sro:eigvec_beta}.


\begin{figure}
    \centering
    \includegraphics[width=.6\linewidth]{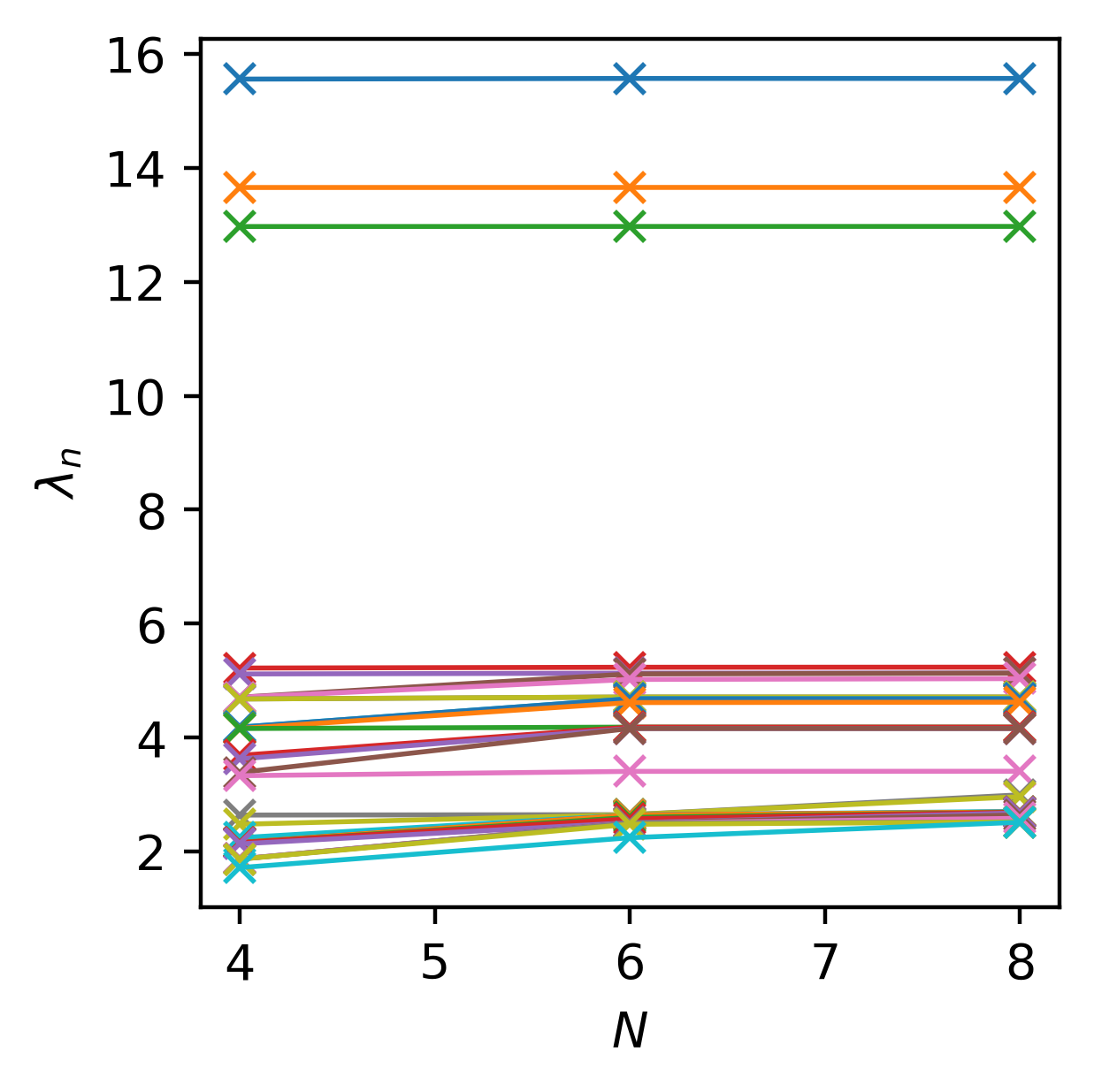}
    \caption{
    \label{fig:sro_conv_N}
    Eigenvalue convergence with respect to the Legendre basis size $N$ at $\beta=80$ and $\phi_{\alpha;\mu_1\mu_2}=0.0002$ for all $\alpha, \mu_1, \mu_2$.
    The three largest eigenvalues are converged at $N=4$.
    At larger $N$, new states appear corresponding to higher frequency version of dominant eigenvectors, discussed in the next section. They can only be captured by considering higher order polynomials.
    At $N=8$, the important eigenvalues are well converged.
    }
\end{figure}

\begin{figure*}
    \centering
    \includegraphics[width=1\linewidth]{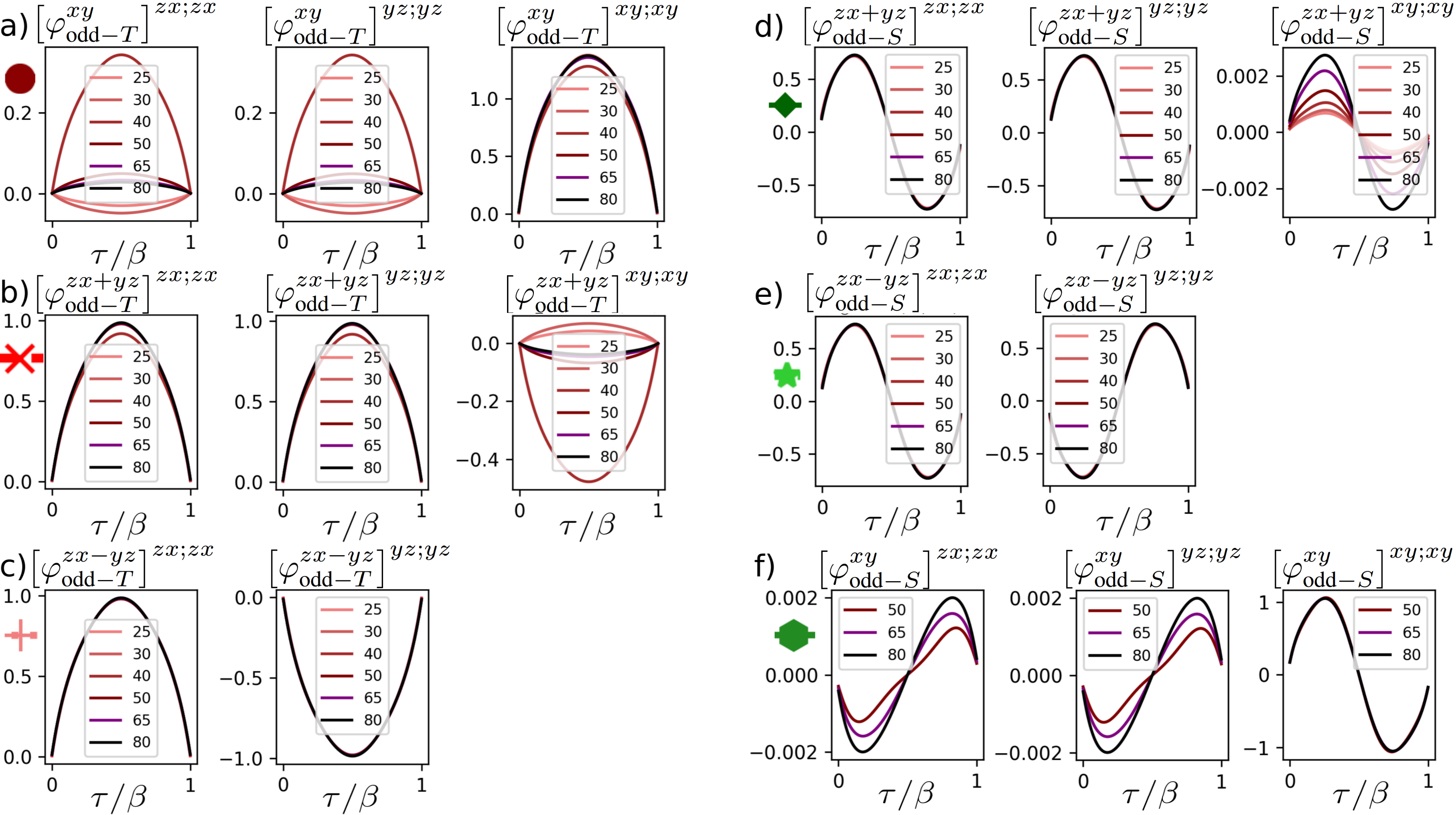}
    \caption{
    Inverse temperature dependence and imaginary-time structure of the non-vanishing spin-orbital components of
    $\varphi^{xy}_{\text{odd}-T}$ in (a), 
    $\varphi^{zx+yz}_{\text{odd}-T}$ in (b) and 
    $\varphi^{zx-yz}_{\text{odd}-T}$ in (c), labelled as $^+S^+P^+O^-T$, along with
    $\varphi^{zx+yz}_{\text{odd}-S}$ in (d),  
    $\varphi^{zx-yz}_{\text{odd}-S}$ in (e), and
    $\varphi^{xy}_{\text{odd}-S}$ in (f), labelled as $^-S^+P^+O^+T$.
    The eigenvalues of these states are shown in \fref{fig:sro_eigval_vs_T}.
    }
    \label{fig:sro_leadvec}
\end{figure*}

\subsection{\label{sec:additional_sro:basis_size_conv} Basis size convergence.}
In \fref{fig:sro_conv_N}, we present the convergence of the eigenvalues as a function of the size of the Legendre basis $N$ at $\beta=80$.
The three largest eigenvalues are converged at $N=4$, since they are simple functions in Legendre-space.
However, including more eigenvalues also includes eigenvectors that have a higher complexity in their imaginary-time structures.
Those eigenstates require a larger number of Legendre polynomials to be well represented.
At $N=8$, the ones presented in this figure are converged for these parameters, and we are not interested in eigenstates with smaller eigenvalues.

\begin{figure}[b]
    \centering
    \includegraphics[width=\linewidth]{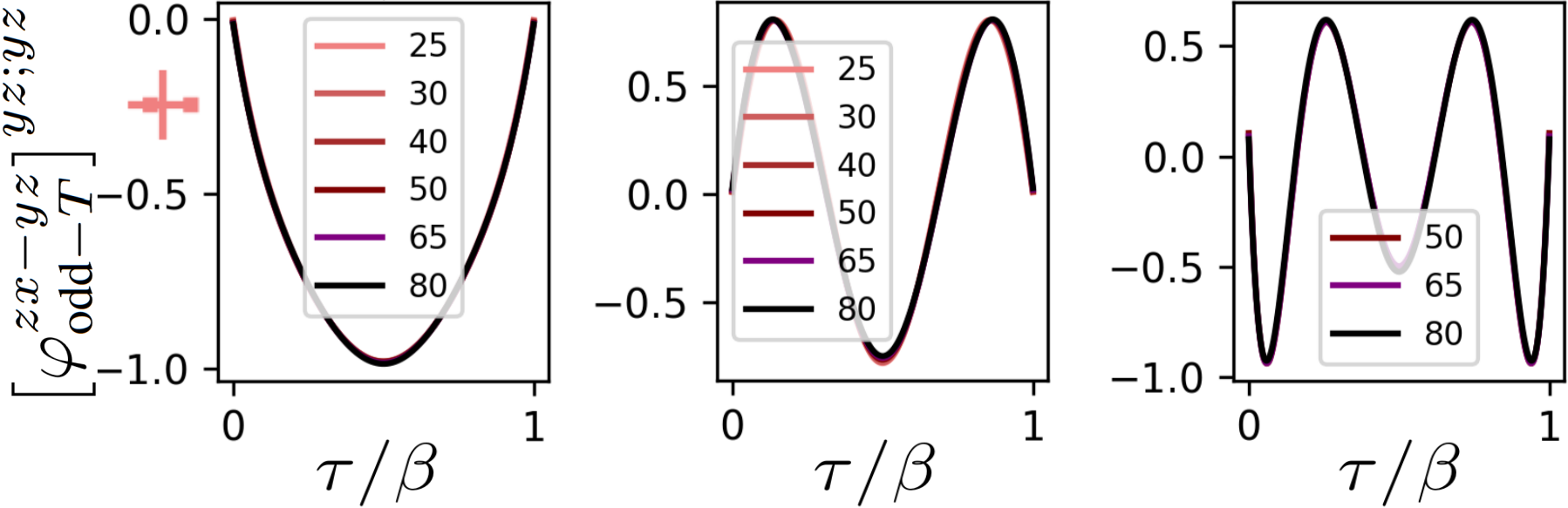}
    \caption{
    Temperature dependence one component ($yz;yz$) of three orthogonal states with the same hosting orbitals and $SPOT$ classification associated to the $\varphi^{zx-yz}_{\text{odd}-T}$ state.
    Once $\tau$ is rescaled by $\beta$, each state preserves it imaginary-time structure with temperature.
    }
    \label{fig:sro_eigvec_harm}
\end{figure}

\subsection{\label{sec:additional_sro:higher-order} Time-higher-order eigenvectors.}
By inspecting \fref{fig:sro_eigval_vs_T}, one will see that the same symmetry and hosting orbitals can appear multiple times in the figure.
The reason is that there can be an infinite number of eigenvectors that share these characteristics, as long as they are orthogonal in imaginary-time.
We refer to those as higher-order in time.

As an example, we present the $yz;yz$ component of the states with $\varphi^{zx-yz}_{\text{odd}-T}$ symmetry for various temperature in \fref{fig:sro_eigvec_harm}.
The imaginary-time structure of these eigenstates is found to be stable with temperature when the imaginary-time axis is rescaled by the inverse temperature.
Moreover, we find that the eigenvalue is reducing with the order of these time-higher-order eigenstates.
In other words, for someone interested in the higher eigenvalues, it appears to be sufficient to consider a smaller Legendre basis, since the largest eigenvalues are well represented by it.
Also, in the cases considered here, the simpler imaginary-time structures are more favorable than the time-higher-order ones.

\subsection{\label{sec:additional_sro:eigvec_beta} Temperature evolution of eigenvectors.}

Finally, \fref{fig:sro_leadvec} presents more detail about the temperature dependence of spin-orbital and imaginary-time structures of the leading eigenvectors.
We only show the non-vanishing spin-orbital components.
Figures~(a), (b) and (c) correspond to the intra-orbital odd-frequency eigenvectors, respectively labelled $\varphi^{xy}_{\text{odd}-T}$, $\varphi^{zx+yz}_{\text{odd}-T}$ and $\varphi^{zx-yz}_{\text{odd}-T}$, while figures~(d), (e) and (f) correspond to the intra-orbital even-frequency eigenvectors, respectively labelled $\varphi^{zx+yz}_{\text{odd}-S}$, $\varphi^{zx-yz}_{\text{odd}-S}$ and $\varphi^{xy}_{\text{odd}-S}$.

In the case of the odd-frequency solutions, decreasing temperature (increasing $\beta$) makes the dominant spin-orbital hosts become dominant, although this components doesn't change too much with temperature.
In the case of the even-frequency solutions, the subdominant spin-orbital hosts become increasingly important with decreasing temperatures, although they remain modest.
The dominant component does not change much with temperature.

As discussed in Ref.~\citenum{gingras_superconductivity_2022}, including spin-orbit coupling would lead to a lot more mixing between the dominant and subdominant hosts.
Further work is necessary to study this point.

\end{document}